\documentclass{aa}
\usepackage{graphicx}
\usepackage{amsmath}
\usepackage{amsfonts}
\usepackage{amssymb}
\usepackage{bm}
\usepackage{natbib}
\usepackage{ulem}
\bibpunct{(}{)}{;}{a}{}{,}
\usepackage{txfonts}
\usepackage{color}
\def\msol{M_\odot}
\def\dst{\displaystyle}
\def\rhog{\rho_{\mathrm{g}}}
\def\rhod{\rho_{\mathrm{d}}}
\def\cs{c_{\mathrm{s}}}
\def\brhog{\bar{\rho}_{\mathrm{g}}}
\def\bcs{\bar{c}_\mathrm{s}}
\def\bp{\bar{P}}
\def\gm{\mathcal{G}M}
\def\vk{v_{\mathrm{k}}}

\def\diffr{\partial_{r}}
\def\diffzp{\partial_{z'}}

\def\Rz{r_{0}}
\def\csz{c_{\mathrm{s}0}}
\def\Tze{T_{0}}
\def\Sigmaz{\Sigma_{0}}
\def\vkz{v_{\mathrm{k}0}}

\def\Hz{H_{0}}
\def\rhogz{\rho_{\mathrm{g}0}}
\def\tsz{t_{\mathrm{s}0}}

\def\csp{c'_{\mathrm{s}0}}
\def\Tp{T_{0}'}
\def\Sigmap{\Sigma_{0}'}
\def\Hp{H_{0}'}
\def\rhogp{\rho_{\mathrm{g}0}'}

\def\dd{z^{2}+r^{2}}

\def\md{m_{\mathrm{d}}}
\def\ts{t_{\mathrm{s}}}
\def\vr{v_{r}}
\def\vtheta{v_{\theta}}
\def\vz{v_{z}}
\def\vgr{v_{\mathrm{g}r}}
\def\vgtheta{v_{\mathrm{g}\theta}}
\def\vgz{v_{\mathrm{g}z}}
\def\tvr{\tilde{v}_{r}}
\def\tvtheta{\tilde{v}_{\theta}}

\def\tkz{t_{\mathrm{k}0}}
\def\vkz{v_{\mathrm{k}0}}
\def\phiz{\phi_{0}}
\def\etaz{\eta_{0}}
\def\etazsq{\eta_{0}^{2}}
\def\sz{S_{0}}
\def\szsq{S_{0}^{2}}

\def\tvrz{\tilde{v}_{r 0}}
\def\tvru{\tilde{v}_{r 1}}
\def\tvrd{\tilde{v}_{r 2}}
\def\tvthetaz{\tilde{v}_{\theta 0}}
\def\tvthetau{\tilde{v}_{\theta 1}}
\def\tvthetad{\tilde{v}_{\theta 2}}

\def\Tm{T_{\mathrm{m}}}

\def\sqrtexpr{\sqrt{\frac{1}{R}-\etaz R^{-q}}}

\def\Tur{T_{\mathrm{A}}\left(R\right)}
\def\Tdr{T_{\mathrm{B}}\left(R\right)}
\def\sm{S_{\mathrm{m}}}

\begin{document}

\title{Revisiting the ``radial-drift barrier'' of planet formation and its relevance in observed protoplanetary discs}

\author{G. Laibe\inst{1,2} \and J.-F. Gonzalez\inst{1} \and S.T. Maddison\inst{3}}


\institute{Universit\'e de Lyon, Lyon, F-69003, France;
Universit\'e Lyon 1, Villeurbanne, F-69622, France;
CNRS, UMR 5574, Centre de Recherche Astrophysique de Lyon;
\'Ecole normale sup\'erieure de Lyon, 46, all\'ee d'Italie,
F-69364 Lyon Cedex 07, France\\
\email{[Guillaume.Laibe;Jean-Francois.Gonzalez]@ens-lyon.fr}
\and
Centre for Stellar and Planetary Astrophysics,
School of Mathematical Sciences, Monash University, Clayton Vic 3168, Australia\\
\email{guillaume.laibe@monash.edu}
\and
Centre for Astrophysics and Supercomputing, Swinburne University,
PO Box 218, Hawthorn, VIC 3122, Australia\\
\email{smaddison@swin.edu.au}}

\date{Received 7 July 2010; Accepted ??}

 
\abstract
{To form metre-sized pre-planetesimals in protoplanetary discs, growing grains have to decouple from the gas before they are accreted onto the central star during their phase of fast radial migration and thus overcome the so-called ``radial-drift barrier'' (often inaccurately referred to as the ``metre-size barrier'').}
{To predict the outcome of the radial motion of dust grains in protoplanetary discs whose surface density and temperature follow power-law profiles, with exponent $p$ and $q$ respectively. We investigate both the Epstein and the Stokes drag regimes which govern the motion of the dust.}
{We analytically integrate the equations of motion obtained from perturbation analysis. We compare these results with those from direct numerical integration of the equations of motion. Then, using data from observed discs, we predict the fate of dust grains in real discs.}
{When a dust grain reaches the inner regions of the disc, the acceleration due to the increase of the pressure gradient is counterbalanced by the increase of the gas drag. We find that most grains in the Epstein (resp. the Stokes) regime survive their radial migration if $-p+q+\frac{1}{2} \le 0$ (resp. if $ q \le \frac{2}{3}$). The majority of observed discs satisfies both $-p+q+\frac{1}{2} \le 0$ and $ q \le \frac{2}{3}$: a large fraction of both their small and large grains remain in the disc, for them the radial drift barrier does not exist.}
{}


\maketitle


\section{Introduction}
\label{Sec:Intro}

Much of the information about the gas structure of protoplanetary discs is inferred from the emission by the dust component and an assumed dust-to-gas ratio. Interpretations of recent observations in the (sub)millimetre domain \citep{AWTau2005,AWOph2007,Lommen2007} show that observed discs typically have masses between $10^{-4}$ and $10^{-1}\ \msol$ and a spatial extent of a few hundred AU. Their radial surface density and temperature profiles are approximated by power laws ($\Sigma \propto r^{-p}$, $\mathcal{T} \propto r^{-q}$), whose respective exponents $p$ and $q$ have positive values typically of order unity.

\defcitealias{Weidendust1977}{W77}\defcitealias{Nakagawa1986}{NSH86}
Seminal studies describe the dust motion in protoplanetary discs, which depends strongly on the gas structure. \citet[hereafter W77]{Weidendust1977} and \citet[hereafter NSH86]{Nakagawa1986} demonstrated that dust grains from micron sizes to pre-planetesimals (a few metres in size) experience a radial motion through protoplanetary discs. This motion is called radial drift or migration. Due to its pressure gradient, the gas orbits the central star at a sub-Keplerian velocity. Grains therefore have a differential velocity with respect to the gas. The ensuing drag transfers linear and angular momentum from the dust to the gas. Thus, dust particles can not sustain the Keplerian motion they would have without the presence of gas and as a result migrate toward the central star.

This migration motion depends strongly on the grain size, which sets the magnitude of the drag, as well as the nature of the drag regime. Specifically, as shown by \citetalias{Weidendust1977} and \citetalias{Nakagawa1986}, grains of a critical size pass through the disc in a fraction of the disc lifetime. This catastrophic outcome is called the ``radial-drift barrier'' of planet formation. More precisely, we will adopt the subsequent definition for the ``radial-drift barrier'' in this study: ``the ability of grains of be accreted onto the central star/depleted from the disc within its lifetime''. Historically, this process was first studied in a Minimum Mass Solar Nebula \citep[MMSN, see][]{WeidenMMSN1977,Hayashi1981,Desch2007,Crida2009}, in which the critical size corresponds to metre-sized bodies and thus was called the ``metre-size barrier''. However, planets are frequently observed (besides the 8 planets in our solar system, more than 700 extra-solar planets have been discovered to date\footnote{http://exoplanet.eu}): some solid material must therefore have overcome this barrier and stayed in the disc to form larger bodies. Moreover, if the small grains of every disc were submitted to the radial-drift barrier, we would barely detect them since their emission via optical/IR scattering and IR thermal radiation is due to small grains. As discs are frequently observed, the grains from which the emission is detected cannot be strongly depleted for a substantial fraction of discs.

\defcitealias{YS2002}{YS02}
From a theoretical point of view, such a discrepancy between the observations and the theoretical predictions imply that the seminal theory has to be extended (some physical element is lacking) or that it has not been fully exploited. This second option has been investigated by \citet[hereafter YS02]{YS2002}. They highlight the fact that, contrary to the primary hypothesis of \citetalias{Weidendust1977}, observed dusty discs are drastically different from the MMSN prototype. As the radial surface density and temperature profiles fix both the radial pressure gradient and the magnitude of the gas drag, different values for the power-law exponents $p$ and $q$ affect the optimal grain size of migration and thus induce different radial motions for the dust through the disc.  Specifically, YS02 showed that for steep surface density profiles and smooth temperature profiles, the grains radial velocity decreases when the grains reach the inner discs regions. Grains in such discs therefore experience a ``pile-up''. However, while important, the work of \citetalias{YS2002} does not  provide a precise conclusion on the outcome of the grains nor any quantitative criterion for the ``pile-up'' process to be efficient enough to avoid the radial-drift barrier. Furthermore, \citetalias{YS2002} restricts their study to the special case of a gas phase with a low density (e.g. the grain size smaller than the gas mean free path, called the Epstein regime). This hypothesis is not valid anymore when considering the radial drift of pre-planetesimals, whose grain sizes are larger than the gas mean free path and are submitted to the Stokes drag regime. Although the radial drift of pre-planetesimals has already been studied in different situations with  numerical or semi-analytical methods --- see e.g. \citet{Haghighipour2003,Birnstiel2009,Youdin2011} --- its rigorous theory for the standard case of a simple disc has not yet been derived.

Within this context, we see that (i) the seminal theory describing the radial motion of dust grains has  been developed within the limits of the Epstein regime but does not treat the Stokes regime ; (ii) here exists no clear theoretical criterion to predict the impact of the ``pile-up effect'' on the outcome of the dust radial motion ; (iii) there exists no criterion to predict whether a given disc will be submitted to the ``radial-drift barrier'' phenomenon. To answer these three points, we re-visit in this study the work of \citetalias{Weidendust1977} and \citetalias{Nakagawa1986} and extend the developments of \citetalias{YS2002} for both the Epstein and the Stokes regime. Performing rigorous perturbative expansions, we find two theoretical criteria (one for each regime) which predict when the ``pile-up'' effect is sufficient for the grains not to be accreted onto the central star. We then test when these theoretical criteria can be applied in real discs.

Additionally, our work is motivated by the recent observational results of \citet{Ricci2010a,Ricci2010b}. From their observations they claim that ``a mechanism halting or slowing down the inward radial drift of solid particles is required to explain the data''.  In this work we aim to show that contrary to what is usually invoked, local pressure maxima due to turbulent vortices or spiral density waves may help but are not necessarily required to explain the observations. \citet{Ricci2010a} also mention that ``the observed flux of the fainter discs are instead typically overpredicted even by more than an order of magnitude''. Here, we also aim to provide a quantitative criterion to determine which discs are faint and which one are not. Thus, revisiting the seminal theory of the radial drift is timely, all the more so than an important quantity of new data is soon to be provided by ALMA, the Atacama Large Millimeter/submillimeter Array.

In this paper, we first recall some general properties of grain motion in protoplanetary discs for both the Epstein and Stokes regime in Sect.~\ref{Sec:GrainDyn}. We then focus on the radial motion of non-growing grains in the Epstein regime. We expand the radial motion equations assuming a weak pressure gradient in Sect.~\ref{Sec:RadialMotion} and detail the two different modes of migration which grains may experience in Sects.~\ref{Sec:Amode} and \ref{Sec:Bmode}. This allows us to derive an analytic criterion which determines the asymptotic dust behaviour in the Epstein regime in Sect.~\ref{Sec:Asymptotic}. We transpose these derivations for the Stokes regime at low Reynolds number in Sect.~\ref{Sec:Stokes} and obtain a  similar criterion for this regime. We also discuss the grains outcome for large Reynolds numbers. In Sect.~\ref{Sec:Discuss}, we discuss the relevance of these criteria and study their implications for observed protoplanetary discs and planet formation in Sect.~\ref{Sec:PlanetFormCsq}. Our conclusions are presented in Sect.~\ref{Sec:Conclusion}.

\section{Dynamics of dust grains}
\label{Sec:GrainDyn}

To reduce the parameter space for this study, we assume the following:

\begin{enumerate}

\item The disc is a thin, non-magnetic, non-self-graviting, inviscid perfect gas disc which is vertically isothermal. Its radial surface density and temperature are described by power-law profiles. Notations are described in Appendix \ref{App:Notations}. The flow is laminar and in stationary equilibrium. Consequently, the gas velocity and density are described by well-known relations, which we present in Appendix \ref{App:discStructure}.

\item The grains are compact homogeneous spheres of fixed radius. The collisions between grains and the collective effects due to large dust concentrations are neglected. When the grains are small compared to the mean free path of the gas ($\lambda_\mathrm{g}>{4s}/{9}$, where $s$ is the grain size), their interactions with the gas are treated by the Epstein drag force for diluted media \citep{Epstein1924,Baines1965,Stepinski1996}. This drag is caused by the transfer of momentum by individual collisions with gas molecules at the grains surface. Assuming specular reflections on the grain and when the differential velocity with the gas is negligible compared to the gas sound speed, the now common expression of the drag force is
\begin{equation}
\left\lbrace
\begin{array}{rcl}
\mathbf{F}_\mathrm{D} & = & - \dst\frac{m_\mathrm{d}}{\ts}\,\Delta\mathbf{v} \\[1em]
t_{\mathrm{s}} & = & \dst\frac{\rho_{\mathrm{d}} s}{\rho_{\mathrm{g}} c_{\mathrm{s}}} \, ,
\end{array}
\right. 
\label{Epsts}
\end{equation}
where $m_\mathrm{d}$ is the dust grain's mass, $t_{\mathrm{s}}$ the stopping time, $\rho_{\mathrm{g}}$ the gas density, $c_{\mathrm{s}}$ the local gas sound speed, $\rho_{\mathrm{d}}$ the intrinsic dust density, and $\Delta\mathbf{v} = \mathbf{v} - \mathbf{v}_{\mathrm{g}}$ the differential velocity between dust and the mean gas motion. In classical T Tauri star (CTTS) protoplanetary discs, drag forces for particles smaller than $\sim 10$~m are well described by the Epstein regime \citep[see also Sect.~\ref{Sec:RealDiscs}]{Garaud2004}. Small grains which produce the emission of observed protoplanetary discs satisfy this criterion.

The interactions between large dust particles ($\lambda_\mathrm{g}<{4s}/{9}$)  and the gas are treated by the Stokes drag force \citep{Whipple1972, Stepinski1996}. In this case, the gas mean free path is small and the dust particle is locally surrounded by a viscous fluid. Depending on the local Reynolds number of the flow around the grains $R_{\mathrm{g}}~=~\frac{2s |\Delta\mathbf{v}|}{\nu}$, where $\nu$ is the microscopic kinematic viscosity of the gas, the drag force takes the following expression:
\begin{equation}
\textbf{F}_{\mathrm{D}} = -\frac{1}{2} C_{\mathrm{D}}  \pi s^{2} \rhog \left| \Delta\mathbf{v}\right| \Delta\mathbf{v},
\label{Stokes}
\end{equation}
where the drag coefficient $C_{\mathrm{D}}$ is given by
\begin{equation}
C_{\mathrm{D}} =
\left\lbrace
\begin{array}{ll}
24 R_{\mathrm{g}}^{-1} & \mathrm{for}\ R_{\mathrm{g}} < 1 \\[1em]
24 R_{\mathrm{g}}^{-0.6} & \mathrm{for}\ 1 < R_{\mathrm{g}} < 800 \\[1em]
0.44 & \mathrm{for}\ 800 < R_{\mathrm{g}} \, .
\end{array}
\right.
\label{Cd_Stokes}
\end{equation}
If $R_{\mathrm{g}} < 1$, the drag force remains linear in $\Delta\mathbf{v}$.
\end{enumerate} 

In this work, the physical relations are written in cylindrical coordinates ($r$, $\theta$, $z$). The related unit vector system is given by $\left( \textbf{e}_{r}, \textbf{e}_{\theta}, \textbf{e}_{z} \right)$. As the system is invariant by rotation around the vertical axis $\textbf{e}_{z}$, the physical quantities depend only on $r$ and $z$. The physical quantities of the gas, designated by subscript g, are first determined in a general way. Then, the limit $z = 0$ is taken to study the restricted radial motion.

Dust dynamics depends on both the magnitude of the drag (driven by the differential velocity) and on its relative contribution with respect to the gravity of the central star. Seminal studies of dust dynamics were conducted by \citet{Whipple1972}, \citetalias{Weidendust1977}, \citet{Weiden1980} and \citetalias{Nakagawa1986}, and extended by others ({\citetalias{YS2002}}; \citealp{Takeuchi2002,Haghighipour2003,Garaud2004,YoudinChiang2004}). Here we recall the major points of those studies. We consider two forces acting on the grain: the gravity of the central star and gas drag. (We assume that the momentum transferred by drag from a single grain on the gas phase is negligible.) Thus

\begin{equation}
\md \frac{\mathrm{d}\textbf{v}}{\mathrm{d}t} = - \textbf{F}_{\mathrm{D}} + \md \textbf{g} ,
\label{pfd_gene}
\end{equation}
where $\textbf{F}_{\mathrm{D}}$ is the drag force. As shown by Eqs.~(\ref{Epsts})--(\ref{Stokes}), a general expression of the ratio $\frac{ \textbf{F}_{\mathrm{D}} }{ m_{\mathrm{d}}  }$ is of the form
\begin{equation}
\frac{ \textbf{F}_{\mathrm{D}} }{ m_{\mathrm{d}}  } = - \frac{\tilde{\mathcal{C}}\left(r,z \right)}{ s^{y}}  | \textbf{v}  - \textbf{v}_{\mathrm{g}} |^{\lambda} \left(\textbf{v}  - \textbf{v}_{\mathrm{g}} \right),
\label{eq:deffdrag}
\end{equation}
where the quantities $\tilde{\mathcal{C}}$, $y$ and $\lambda$ are defined for both the Epstein and the Stokes regime in Appendix \ref{App:Dimensionless}.

%
\section{Radial motion in the Epstein regime: perturbation analysis at small pressure gradients}
\label{Sec:RadialMotion}

Considering the Epstein (small grains) regime, Eq. (\ref{pfd_gene}) reduces to:
\begin{equation}
\md \frac{\mathrm{d}\textbf{v}}{\mathrm{d}t} = - \frac{\md}{\ts} \left(\textbf{v} - \textbf{v}_{\mathrm{g}} \right) + \md \textbf{g} .
\label{pfd}
\end{equation}
Writing Eq.~(\ref{pfd}) in $\left(r,\theta ,z \right)$ coordinates leads to
\begin{equation}
\left\lbrace 
\begin{array}{rcl}
\dst \frac{\mathrm{d} \vr}{\mathrm{d} t} - \frac{\vtheta^{2}}{r} + \frac{\left(\vr - \vgr \right)}{\ts} + \frac{\gm r}{\left(\dd \right)^{3/2}} & = & \dst 0 \\
\dst \frac{\mathrm{d} \vtheta}{\mathrm{d} t} + \frac{\vr \vtheta}{r} + \frac{\left(\vtheta - \vgtheta \right)}{\ts} & = & \dst 0 \\
\dst \frac{\mathrm{d} \vz}{\mathrm{d} t} + \frac{\left(\vz - \vgz \right)}{\ts} + \frac{\gm z}{\left(\dd \right)^{3/2}}& = & \dst 0 .
\end{array}
\right.
\label{pdfproj}
\end{equation}
To highlight the important parameters involved in the grains dynamics, we introduce dimensionless quantities (see Appendix \ref{App:Dimensionless}). It is crucial to note that the ratio $\frac{\ts}{t_{\mathrm{k}}}$ of the two timescales related to the physical processes acting on the grain is given by

\begin{equation}
\frac{\ts}{t_{\mathrm{k}}} = \frac{\tsz}{\tkz} R^{p}\mathrm{e}^{\frac{Z^{2}}{2R^{3-q}}} = \sz R^{p}\mathrm{e}^{\frac{Z^{2}}{2R^{3-q}}}.
\label{defsz}
\end{equation}
With Eq.~(\ref{Epsts}) and noting $\Omega_{\mathrm{k}}$ the Keplerian angular velocity, this ratio can be written as
\begin{equation}
\frac{\ts}{t_{\mathrm{k}}} = \frac{s}{\left(\frac{\rhog \cs}{\rhod \Omega_{\mathrm{k}}}\right)} = \frac{s}{s_{\mathrm{opt}}} = S ,
\label{defssopt}
\end{equation}
where $s_{\mathrm{opt}} = \frac{\rhog \cs}{\rhod \Omega_{\mathrm{k}}}$. This timescale ratio therefore corresponds to a dimensionless size $S = \sz R^{p}e^{\frac{Z^{2}}{2R^{3-q}}}$ for the grain. If $S \ll 1$ (resp. $S \gg 1$), the effects of drag will occur much faster (resp. slower) than gravitational effects. If $S \simeq 1$, both gravity and drag will act on the same timescale. Interestingly, $s_{\mathrm{opt}}$ varies in the disc midplane as $r^{-p}$, as does surface density.

Then using Eqs.~(\ref{defsz}), (\ref{vgasnul}) and (\ref{dimensionlessgas}), we obtain for ($\textbf{e}_{r},\textbf{e}_{\theta},\textbf{e}_{z}$):
\begin{equation}
\left\lbrace 
\begin{array}{rcl}
\dst \frac{\mathrm{d} \tvr}{\mathrm{d} T} - \frac{\tvtheta^{2}}{R} + \frac{\tvr}{\sz}R^{-\left(p+\frac{3}{2} \right)}\mathrm{e}^{-\frac{Z^{2}}{2R^{3-q}}} +\frac{R}{\left(R^2 + \phiz Z^{2} \right)^{3/2}} & = & 0 \\[2ex]
\multicolumn{1}{l}{\dst \frac{\mathrm{d} \tvtheta}{\mathrm{d} T} + \frac{\tvtheta\tvr}{R} +} &&\\
\frac{\left(\tvtheta - \sqrt{\frac{1}{R} - \etaz R^{-q} - q\left(\frac{1}{R} - \frac{1}{\sqrt{R^{2}+\phiz Z^{2}}} \right)}\right)}{\sz}R^{-\left(p+\frac{3}{2} \right)}\mathrm{e}^{-\frac{Z^{2}}{2R^{3-q}}} & = & 0 \\[2ex]
\dst \frac{\mathrm{d}^{2}Z}{\mathrm{d}T^{2}} + \frac{1}{\sz}\frac{\mathrm{d}Z}{\mathrm{d}T}R^{-\left(p+\frac{3}{2} \right)}\mathrm{e}^{-\frac{Z^{2}}{2R^{3-q}}} + \frac{Z}{\left(R^{2} + \phiz Z^{2} \right)^{3/2}} & = & 0 .
\end{array}
\right.
\label{dust3d}
\end{equation}
These equations depend on five control parameters: the initial dimensionless grain size, $\sz$, the radial surface density and temperature exponents, $p$ and $q$, the square of the disc aspect ratio, $\phiz = \left(\Hz/\Rz \right)^{2}$, and the subkeplerian parameter, $\etaz$, given by Eq.~(\ref{defetaz}). The equations can be simplified in some cases, e.g. if the vertical motion is considered to occur faster than the radial motion, $R\simeq1$ and $\frac{\mathrm{d}^{2}Z}{\mathrm{d}T^{2}}$ simplifies to the damped harmonic oscillator equation. If we consider only the radial motion (for a 2D disc), we have $Z=0$, and
\begin{equation}
\left\lbrace
\begin{array}{l}
\dst \frac{\mathrm{d} \tvr}{\mathrm{d} T}  =  \dst \frac{\tvtheta^{2}}{R} - \frac{\tvr}{\sz}R^{-\left(p+\frac{3}{2} \right)} -\frac{1}{R^2} \\
\dst \frac{\mathrm{d} \tvtheta}{\mathrm{d} T}  =  \dst -\frac{\tvtheta \tvr}{R} - \frac{\left(\tvtheta - \sqrt{\frac{1}{R} - \etaz R^{-q}}\right)}{\sz}R^{-\left(p+\frac{3}{2} \right)} .
\end{array}
\right.
\label{radialseulmodif}
\end{equation}

\begin{figure}
\resizebox{\hsize}{!}{\includegraphics{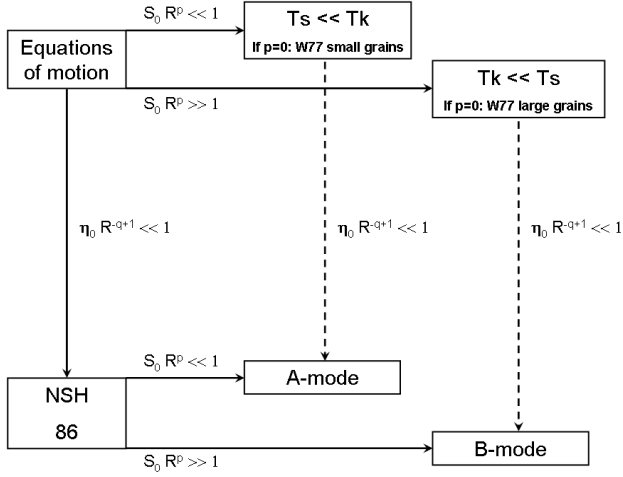}}
\caption{Principle of the various perturbative expansions for the grain radial motion. Expanding first with respect to the small pressure gradient ($\etaz R^{-q+1} \ll 1$) leads to NSH86 equations. Expanding first with respect to the grain sizes ($\sz R^{p} \ll 1$ or $\sz R^{p} \gg 1$) leads to W77 expressions for the particular case $p = 0$. Combining both leads to  the A- and B- mode, respectively for small and large grains.}
\label{expandshem}
\end{figure}

Even for discs in two dimensions, Eq.~(\ref{radialseulmodif}) is not analytically tractable. However, as some of the parameters involved in the equation are small, approximations of the solution can be found by performing perturbative expansions. Some of the classical results detailed below have been studied in \citetalias{Weidendust1977} and \citetalias{Nakagawa1986}, but are here properly justified. The principle of those expansions is described on Fig.~\ref{expandshem}.
At $t = 0$, $r = \Rz$ which implies that $R\left(T = 0 \right) = 1$. Because of gas drag, a grain dissipates both its energy and angular momentum and therefore, experiences a radial inward motion, i.e. $R<1$. The first parameter with respect to which a perturbative expansion can be performed is $\etaz$ (linked to the pressure gradient by Eq.~(\ref{eqjustufyW77})) as it takes values of approximately $10^{-3}$--$10^{-2}$ in real protoplanetary discs (see \citetalias{Nakagawa1986}), and thus $\etaz \ll 1$. We consider that this inequality also implies that
\begin{equation}
\etaz R^{-q} \ll \frac{1}{R} .
\label{ineqterm}
\end{equation}

This inequality is always verified when $q\le1$ and thus applies to observed discs (see Sect.~\ref{Sec:PlanetFormCsq}). For $q>1$, there is a region where this inequality is not verified. However, in this case, the pressure gradient has the same order of magnitude as the gravity of the central star and the model of a power-law profile for the radial temperature is not accurate enough to model realistic discs. We thus consider that for real discs Eq.~(\ref{ineqterm}) is always justified. Then, following \citetalias{Nakagawa1986}, we consider the system of equations given by Eq.~(\ref{radialseulmodif}). We set
\begin{equation}
\left\lbrace
\begin{array}{l}
\dst \tvr  =  \dst \tvrz + \etaz \tvru + \mathcal{O}\left(\etaz^{2}\right)\\
\dst \tvtheta  = \dst \tvthetaz + \etaz \tvthetau + \mathcal{O}\left(\etaz^{2}\right)
\end{array}
\right.
\label{dvptnaka}
\end{equation}
and look at the orders $\mathcal{O}\left(1\right)$, $\mathcal{O}\left(\etaz\right)$,... of the expansion -- see Appendix \ref{App:Epstein_perturb}. We find that:
\begin{equation}
\tvr = \etaz \tvru + \mathcal{O}\left(\etazsq \right) = -\frac{2 \sz R^{p- \frac{1}{2} } \left(1 - \sqrt{1 - \etaz R^{-q+1}} \right)}{1 + R^{2p} \szsq} + \mathcal{O}\left(\etaz ^{2} \right). 
\label{nakatvru}
\end{equation}
The pressure gradient term has been retained to keep the generality, however since we assume that $\etaz \ll 1$, we equivalently have
\begin{equation}
\sqrt{\frac{1}{R} - \etaz R^{-q}} - \sqrt{\frac{1}{R}} = -\frac{\etaz}{2} R^{-q+\frac{1}{2}} + \mathcal{O}\left(\etaz ^{2} \right) .
\label{approxsmalletaz}
\end{equation}
Thus, to order $\mathcal{O}\left(\etaz \right)$,
\begin{equation}
\tvr = -\frac{\etaz \sz R^{p-q+\frac{1}{2}} }{1 + R^{2p} \szsq} + \mathcal{O}\left(\etazsq \right),
\label{nakatvrusimp}
\end{equation}
or equivalently, using Eqs.~(\ref{dimensionlessdust}) and (\ref{dimensionlessgas}),
\begin{equation}
\vr = \frac{r c_{\mathrm{s}}^{2}}{\vk} \frac{\mathrm{d} \, \mathrm{ln} \bp}{\mathrm{d} r} \frac{\left(\ts /t_{\mathrm{k}} \right)}{1+\left(\ts / t_{\mathrm{k}} \right)^{2}} .
\label{nakatvrusimpdev}
\end{equation}
$\sz R^{p}$ is the dimensionless expression of the ratio $\ts/t_{\mathrm{k}}$. Eq.~(\ref{nakatvrusimp}) shows that $R^{2p}\szsq \ll 1$ or $R^{2p}\szsq \gg 1$, and thus $R^{p}\sz \ll 1$ or $R^{p}\sz \gg 1$, resulting in asymptotic behaviours for the radial grain motion. These asymptotic regimes were first described by \citetalias{Weidendust1977} for the particular case $p = 0$. They correspond physically to two limiting cases: where the gas drag dominates, which we call the A-mode, and where gravity dominates, which we call the B-mode. In the next sections, we study and describe these two so-called ``regimes of migration'' or ``modes of migration'' before treating the global evolution of grains given by Eq.~(\ref{nakatvrusimp}). 
%
\subsection{A-mode (Radial differential migration)}
\label{Sec:Amode}

The A-mode corresponds to the regime $R^{p}\sz \ll 1$ (or equivalently $\ts/t_{\mathrm{k}} \ll 1$). In the A-mode, Eq.~(\ref{nakatvrusimp}) reduces to
\begin{equation}
\tvr = \frac{\mathrm{d} r}{\mathrm{d} t} = \frac{\mathrm{d} R}{\mathrm{d} T} = -\etaz \sz R^{p-q+\frac{1}{2}} ,
\label{radmodeA}
\end{equation}
or equivalently
\begin{equation}
\vr = \frac{r c_{\mathrm{s}}^{2}}{\vk} \frac{\mathrm{d} \, \mathrm{ln} \bp}{\mathrm{d} r} \frac{\ts}{t_{\mathrm{k}}} ,
\label{radmodeAdev}
\end{equation}
where the $\mathcal{O}\left(R^{p}\sz \right)$ has been neglected. In this mode of migration, the stopping time is much smaller than the Keplerian time scale. Considering one grain's orbit around the central star, its orbital velocity is forced by the gas drag to become sub-Keplerian in just a few stopping times, i.e. almost instantaneously. Thus, the centrifugal acceleration is not efficient enough to counterbalance the gravitational attraction of the central star, and the grain feels an inward radial differential acceleration. The gas drag counterbalances this radial motion and the grain reaches a local limit velocity in a few stopping times. We call the physical process of the A-mode of migration ``Radial Differential Migration''.

The A-mode of migration originates first from a perturbative expansion for $\etaz \ll 1$ (rigorously for $\etaz R^{-q+1} \ll 1$) and second from a perturbative expansion for $\sz \ll 1$ (rigorously for $\sz R^{p} \ll 1$). Formally, we have performed: $\lim\limits_{\substack{\sz \ll 1}} \lim\limits_{\substack{\etaz \ll 1}}\left[... \right]$. Historically, the A-mode had been derived by \citetalias{Weidendust1977} to explain the radial motion of small grains. In his study, he neglected the radial dependence of the stopping time and assumed $\sz \ll 1$ (this approximation also implies that $R^{2p}\szsq \ll 1$, as $R < 1$, see Appendix \ref{App:W77}).

It is straightforward to integrate the differential equation Eq.~(\ref{radmodeA}) by separating the $R$ and $T$ variables. Noting that $R\left(T = 0 \right)= 1$, we have:
\begin{itemize}
\item If $-p+q+\frac{1}{2} \ne 0$:
\begin{equation}
\left\lbrace
\begin{array}{l}
\dst R  =  \dst \left[1 - \left(-p+q+\frac{1}{2} \right)\etaz \sz T \right]^{\frac{1}{-p+q+\frac{1}{2}}} \\
\dst T  =  \dst \frac{1 - R^{-p+q+\frac{1}{2}}}{\left(-p+q+\frac{1}{2} \right)\etaz \sz} .
\end{array}
\right.
\label{modeAeq}
\end{equation}

\item If $-p+q+\frac{1}{2} = 0$:
\begin{equation}
\left\lbrace
\begin{array}{l}
\dst R  =  \dst \mathrm{e}^{-\etaz \sz T} \\
\dst T  =  \dst -\frac{\mathrm{ln}\left(R \right)}{\etaz \sz} .
\end{array}
\right.
\label{modeAneq}
\end{equation}

\end{itemize}

The outcome of the dust radial motion comes from a competition between two effects. As the grain reaches smaller radii, (1) gas drag increases, which slows down the radial drift, and (2) the differential acceleration due to the pressure gradient increases which enhances the migration efficiency. Point (1) is related to $s_{\mathrm{opt}}$, which scales as the surface density profile, while the acceleration due to the pressure gradient in (2) is related to the temperature profile (see Eq. (\ref{dimensionlessgas})). Depending on which process is dominant, the grain's dynamics can lead to two regimes:

\begin{itemize}

\item If $-p+q+\frac{1}{2} \le 0$:
then as
\begin{equation}
T (R)  \sim \dst \frac{- R^{-p+q+\frac{1}{2}}}{\left(-p+q+\frac{1}{2} \right)\etaz \sz} =  \mathcal{O} \left(R^{-p+q+\frac{1}{2}} \right) ,
\label{modeAequiv}
\end{equation}
the time it takes the grain to reach smaller and smaller radii increases drastically, according to the diverging power-law. Importantly, this behaviour constitutes our definition of the grain ``pile-up''. Mathematically speaking, accretion onto the central star occurs in an infinite time, i.e.
\begin{equation}
\lim\limits_{\substack{T \to +\infty}} R = 0 .
\label{modeAinf}
\end{equation}

\item If $-p+q+\frac{1}{2} > 0$:
the grain is accreted onto the central star in a finite migration time given by
\begin{equation}
\Tm = \frac{1}{\etaz \sz \left( -p+q+\frac{1}{2} \right)} ,
\label{deftmssgrowth}
\end{equation}
which increases as $\sz$ and $\etaz$ decrease, so that
\begin{equation}
\lim\limits_{\substack{T \to \Tm}} R = 0 .
\label{modeAninf}
\end{equation}
\end{itemize}
The presence or absence of a physical grain pile-up is therefore demonstrated considering the asymptotic behaviour of $R(T)$ at large times. It is important to realise that the pile-up is a cumulative effect that can not arise from velocities only (which however provide qualitative information on the grain's motion) but can only be found by integrating the equation of motion. This rigorously allows us to distinguish two different behaviours for the outcome of the grain's radial motion, and thus two classes of discs with respect to the A-mode.
%
\subsection{B-mode (Drift forced by a resistive torque)}
\label{Sec:Bmode}

Returning to Eq.~(\ref{nakatvrusimp}), the B-mode corresponds to the other asymptotic regime, where $R^{p}\sz \gg 1$ (or equivalently $\ts/t_{\mathrm{k}} \gg 1$). In this case, 
\begin{equation}
\tvr = \frac{\mathrm{d} R}{\mathrm{d} T} = -\frac{\etaz}{\sz} R^{-p-q+\frac{1}{2}},
\label{radmodeB}
\end{equation}
or equivalently
\begin{equation}
\vr = \frac{r c_{\mathrm{s}}^{2}}{\vk} \frac{\mathrm{d} \, \mathrm{ln} \bp}{\mathrm{d} r} \frac{t_{\mathrm{k}}}{\ts} .
\label{radmodeBdev}
\end{equation}
In this mode of migration, the stopping time is much larger than the Keplerian time scale. Hence, the orbital velocity of a grain around the central star is almost the Keplerian velocity. However, because of the pressure gradient, the gas orbits around the central star at a sub-Keplerian velocity. Thus, the azimuthal differential velocity between the gas and the grain generates an azimuthal drag force whose torque extracts angular momentum from the grain. Given that the Keplerian angular momentum increases with radius ($l \propto \sqrt{r}$), this torque results in the inward migration of the grain. We call the physical process of this B-mode of migration ``Drift Forced by a Resistive Torque''.

As for the A-mode, the B-mode of migration can also be derived first from an expansion in $\left(\sz R^{p} \right)^{-2}$ and then from an expansion in $\etaz$. Historically, \citetalias{Weidendust1977} found an expression while only assuming that $\sz \gg 1$ since he considered a flat density profile.
To find the expression derived by \citetalias{Weidendust1977} for large grains, we must assume that $\sz R^{p} \gg 1$. It is crucial to see that this expression does not imply that $\sz \gg 1$  when $p>0$ and $R \to 0$.

It is straightforward to integrate the differential equation Eq.~(\ref{radmodeB}) by separating the $R$ and $T$ variables. Noting that $R\left(T = 0 \right)= 1$, we have:
\begin{itemize}
\item If $p+q+\frac{1}{2} \ne 0$:
\begin{equation}
\left\lbrace
\begin{array}{l}
\dst R  = \dst \left[1 - \left(p+q+\frac{1}{2} \right)\frac{\etaz}{\sz} T \right]^{\frac{1}{p+q+\frac{1}{2}}} \\
\dst T  = \dst \frac{\sz}{\etaz}\frac{1 - R^{p+q+\frac{1}{2}}}{\left(p+q+\frac{1}{2} \right)} .
\end{array}
\right.
\label{modeBeq}
\end{equation}

\item If $p+q+\frac{1}{2} = 0$:
\begin{equation}
\left\lbrace
\begin{array}{l}
\dst R  =  \dst \mathrm{e}^{-\frac{\etaz}{\sz} T} \\
\dst T  =  \dst -\frac{\sz}{\etaz}\mathrm{ln}\left(R \right) .
\end{array}
\right.
\label{modeBneq}
\end{equation}

\end{itemize}

As for the A-mode, two kinds of behaviours appear, depending on the $p$ and $q$ exponents:

\begin{itemize}

\item If $p+q+\frac{1}{2} \le 0$:
The grain migrates inwards, piles-up in the disc's inner regions and falls  onto the star in an infinite time:
\begin{equation}
\lim\limits_{\substack{T \to +\infty}} R = 0 .
\label{modeBinf}
\end{equation}
However, the negative exponents required to be in this regime do not correspond to physical discs. Therefore, the grain dynamics in the B-mode in real discs belong to the second case: 

\item If $p+q+\frac{1}{2} > 0$:
The grain is accreted onto the central star in a finite time
\begin{equation}
\Tm = \frac{\sz}{\etaz \left(p+q+\frac{1}{2} \right)} ,
\label{tmmodeb}
\end{equation}
which increases as $\sz$ increases and $\etaz$ decreases and so that
\begin{equation}
\lim\limits_{\substack{T \to \Tm}} R = 0 .
\label{modeBninf}
\end{equation}

\end{itemize}

As for the A-mode, considering the limit of $R(T)$ at large times also proves the existence of two classes of discs with respect to the B-mode of migration. The radial motion of a grain in the B-mode of migration is also driven by a competition between the increase of both the drag and the acceleration due to the pressure gradient. However, for real discs, $p>0$ and $q>0$, and therefore, $p+q+\frac{1}{2}>0$. Grains migrating in B-mode in such discs fall in a finite time onto the central star.

%
\subsection{Radial evolution and asymptotic behaviour of single grains}
\label{Sec:Asymptotic}

As we have seen, the grains behaviour is divided into two asymptotic regimes, called the A-mode and the B-mode, which come from two different physical origins. However, the two criteria determining if the grains are accreted onto the central object in a finite or infinite time differ for the A- and the B-mode. It is thus crucial to determine in which mode a dust grain ends its motion to predict if the grain is ultimately accreted or not. Returning again to Eq.~(\ref{nakatvrusimp}), we have
\begin{equation}
\frac{\mathrm{d} R}{\mathrm{d} T} = \frac{-\etaz \sz R^{p-q+\frac{1}{2}}}{1 + R^{2p}\szsq} .
\label{rappelNSH86}
\end{equation}
We can separate the $R$ and $T$ variables and integrate to obtain an expression for $T(R)$:
\begin{equation}
T = \frac{1}{\etaz \sz} \left[ \Tur + \Tdr \right] ,
\label{implicitNSH}
\end{equation}
where
\begin{equation}
\Tur  =
\left\lbrace
\begin{array}{ll}
  \dst\frac{1 - R^{-p+q+\frac{1}{2}} }{-p+q+\frac{1}{2}} & \mathrm{if}\ -p+q+\frac{1}{2} \ne 0 \\[1em]
 -\mathrm{ln}\left(R\right) & \mathrm{if}\ -p+q+\frac{1}{2} = 0.
\end{array} 
 \right.
\label{defNu}
\end{equation}
and
\begin{equation}
\Tdr  = 
\left\lbrace
\begin{array}{ll}
 \dst\szsq \frac{1 - R^{p+q+\frac{1}{2}}}{p+q+\frac{1}{2}} & \mathrm{if}\ p+q+\frac{1}{2} \ne 0\\[1em]
 \dst-\szsq \mathrm{ln}\left(R\right) &\mathrm{if}\ p+q+\frac{1}{2} = 0.
\end{array} 
 \right.
\label{defNd}
\end{equation}
Eq.~(\ref{implicitNSH}) provides the asymptotic behaviour of the grains at large times. Interestingly, as $p \ge 0$ for realistic discs, the contribution of the B-mode becomes negligible when $R \ll S_{0}^{-1/p}$. Hence, grains initially migrating in the A-mode stay in the A-mode, but grains initially migrating in the B-mode end their radial motion in the A-mode. This behaviour is summarized on Fig.~\ref{Modesnormal} and detailed in Appendix \ref{App:Asymptotic_radial}. This result was not predicted by \citetalias{Weidendust1977}, as he neglected the radial dependence of the stopping time. Mathematically speaking, it comes from the fact that the perturbative expansion of \citetalias{Weidendust1977} has been performed with respect to powers of $\sz$ and not powers of $\sz R^{p}$.
\begin{figure}
\resizebox{\hsize}{!}{\includegraphics{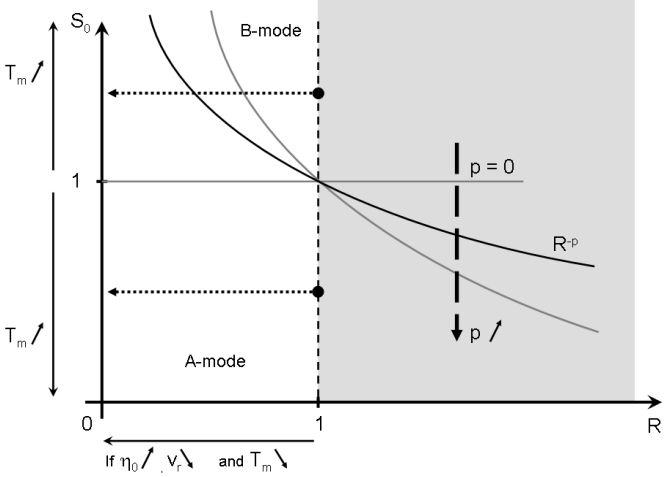}}
\caption{Radial evolution of grains in the ($R$,$\sz$) plane showing that a grain in the Epstein drag regime ends its radial motion in the A-mode.  The solid curves represent $R^{-p}$ for various values of $p$, they separate the A-mode (below) from the B-mode (above) regions. The horizontal dashed lines show trajectories of grains as they migrate inwards from $R = 1$. The shaded area is a forbidden zone.}
\label{Modesnormal}
\end{figure}

To illustrate the radial motion of dust grains in protoplanetary discs, we numerically integrate the equations of motion for different values of the parameters $\etaz$, $\sz$, $p$, $q$. We set $\etaz = 10^{-2}$ to mimic a realistic disc and vary the order of magnitude of $\sz$ from $10^{-4}$ to $10^{2}$ for two sets of ($p$,$q$) values. First, we choose ($p=0$, $q=\frac{3}{4}$); according to the \citetalias{Nakagawa1986} expansion, the grains are accreted by the central star in a time $\Tm$. Second, we set ($p=\frac{3}{2}$, $q=\frac{3}{4}$); the grains fall onto the central star in an infinite time from the same approximation. This set of ($p$,$q$) values is taken to mimic discs profiles that are commonly used and for which $-p+q+\frac{1}{2}$ can take a positive or a negative value. Consequently, we interpret the radial motion $R\left(T\right)$ of dust grains plotted in Fig.~\ref{plotbothcases}.
\begin{itemize}

\item The top panel of Fig.~\ref{plotbothcases} shows the results for ($p=0$, $q=\frac{3}{4}$): Grains fall onto the central star, initially in the A-mode for the small grains and in the B-mode for the large ones. The radial-drift process is long for small and large grains but is optimal for grains with $ S = \sm = 1$ for which the accretion time is $\Tm = 1.6/ \etaz$, or $\Tm=160$ with $\etaz=10^{-2}$.

\begin{figure}
\resizebox{\hsize}{!}{\includegraphics[angle=-90]{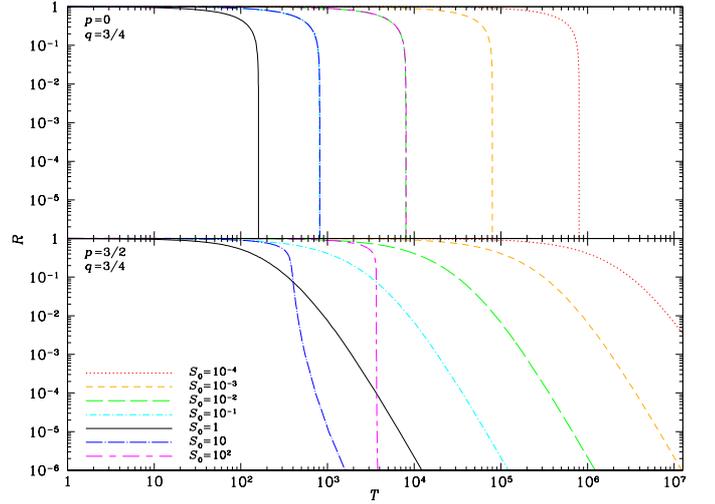}}
\caption{Radial motion $R\left(T\right)$ of dust grains in the Epstein regime for $\etaz = 10^{-2}$. $\sz$ varies from $10^{-4}$ to $10^{2}$. Top: $p=0$, $q=\frac{3}{4}$, here $-p+q+\frac{1}{2}>0$ and the grain is accreted onto the central star in a finite time. Bottom: $p=\frac{3}{2}$, $q=\frac{3}{4}$, here $-p+q+\frac{1}{2}<0$ and the grain piles up and is consequently accreted onto the central star in an infinite time.} 
\label{plotbothcases}
\end{figure}

\item The bottom panel of Fig.~\ref{plotbothcases} shows the results for ($p=\frac{3}{2}$, $q=\frac{3}{4}$): In this case, the radial density profile is steep enough to ensure that the grains are not accreted onto the central star. To reach a given radius (for example $R_{\mathrm{f}} = 0.1)$, the optimal size is $S_{\mathrm{m,f}} \simeq 2.9 = \mathcal{O}\left(1\right)$ (see Eq.~(\ref{exprsmr})). Hence, in this case grains efficiently reach the disc inner regions without ever being accreted onto the central star. The transition from the B-mode to the A-mode (for which $R \propto T^{-4}$ in this case) for the large grains is visible in this plot.

\end{itemize}

\section{\textbf{Radial motion in the Stokes regime}}
\label{Sec:Stokes}

Radial migration of large particles occurs in the Stokes drag regime, which depends on the dynamical viscosity $\mu$ of the gas. For hydrogen molecules:
\begin{equation}
\mu = \frac{5m\sqrt{\pi}}{64 \sigma_{\mathrm{s}} } \sqrt{\frac{k_{\mathrm{B}} T}{m} } ,
\label{ViscoSpheres}
\end{equation}
where $m = 2m_{\mathrm{H}} = 3.347446922\times10^{-27}$~kg and $\sigma_{\mathrm{s}} = 2.367 \times10^{-19}$~m$^{2}$ is the molecular cross section of the molecule \citep{ChapmanCowling}. The kinematic viscosity $ \nu$ is then defined by $\mu = \rhog \nu$ and the gas collisional mean free path is given by
\begin{equation}
\lambda_{g} = \sqrt{\frac{\pi}{2}} \frac{\nu}{\cs} .
\label{def_mfp}
\end{equation}

We now generalise the procedure outlined in Sects.~\ref{Sec:GrainDyn} and \ref{Sec:RadialMotion} to the three Stokes regimes of Eq.~(\ref{Cd_Stokes}). Using the dimensionless coordinates described above, we have
\begin{equation}
\mu = \mu_{0} R^{-\frac{q}{2}},
\label{genemu}
\end{equation}
\begin{equation}
 \mu_{0}  =  \frac{5m\sqrt{\pi}}{64 \sigma_{\mathrm{s}}} \csz ,
\label{genemuspheres}
\end{equation}
where $\csz$ is given by Eq.~(\ref{powerprofiles}). Thus, the expression of the kinematic viscosity $\nu$ is
\begin{equation}
\nu  =  \dst \nu_{0}R^{\frac{3}{2} +p - q} e^{\frac{Z^{2}}{2 R^{3-q} } } ,
\label{genenu}
\end{equation}
\begin{equation}
 \nu_{0}  =  \dst \frac{\mu_{0}}{\rhogz} \, .
\label{genenuz}
\end{equation}
First, if $R_{\mathrm{g}} <1$, the drag force is linear in $\textbf{v} - \textbf{v}_{\mathrm{g}}$ and thus has the same structure as for the Epstein regime. Comparing the expressions of $\mathcal{C}\left(R,0 \right)$ for the Epstein and the linear Stokes regime (see Appendix~\ref{App:Dimensionless}), all the results found for the radial motion in the Epstein regime can therefore be directly transposed by setting $q' = q$ and $p' = \frac{q-3}{2}$. In this case, the grain radial motion does not depend on $p$ anymore and the \citetalias{Nakagawa1986} expansion of the radial motion for small pressure gradients provides (see Eq.~(\ref{rappelNSH86}))
\begin{equation}
\frac{\mathrm{d} R}{\mathrm{d} T} = \frac{-\etaz S_{0}^{2} R^{-\frac{q}{2} - 1}}{1 + R^{q-3} S_{0}^{4}} .
\label{NSH86Stokes}
\end{equation}
These crucial results follow:
\begin{itemize}

\item In the A-mode ($\sz \ll R^{\frac{3 - q}{4}}$), grains experience a pile-up and migrate onto the central star in an infinite time if $-p'+q'+\frac{1}{2}\le0$, i.e.\ if $q \le -4$ (which never occurs in real discs). 

\item In the B-mode ($\sz \gg R^{\frac{3 - q}{4}}$), grains migrate onto the central star in an infinite time if $p'+q'+\frac{1}{2}\le0$, i.e.\ if $q\le\frac{2}{3}$.

\end{itemize}

\begin{figure}
\resizebox{\hsize}{!}{\includegraphics{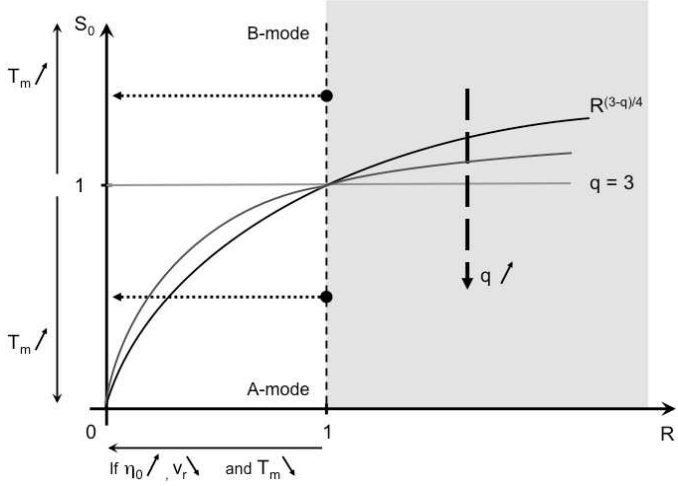}}
\caption{Radial evolution of grains in the ($R$,$\sz$) plane showing that a grain in the linear Stokes drag regime ends its radial motion in the B-mode. The solid curves represent $R^{\frac{3-q}{4}}$ for various values of $q$, they separate the A-mode (below) from the B-mode (above) regions. The horizontal dashed lines show trajectories of grains as they migrate inwards from $R = 1$. Shaded area: forbidden.}
\label{ModesnormalStokes}
\end{figure}
Thus, similar to the Epstein regime, we derived one criterion for each mode and need to determine in which mode the grain ends its motion. For observed discs, $q-3 < 0$ (see Sect.~\ref{Sec:PlanetFormCsq}), and as the particle migrates inward, $R$ becomes smaller than $S_{0}^{\frac{4}{3 - q}}$ and grains end their radial motion in the B-mode (see Fig.~\ref{ModesnormalStokes}). This result is fundamentally different to the one we obtained for the Epstein regime. Indeed, for grains migrating in the A-mode in the Stokes regime at low Reynolds numbers, the criterion obtained for a pile-up in the A-mode is never satisfied for real discs. However, after migrating inside a critical radius, grains switch to the B-mode, for which the pile-up can potentially occur, depending on the value of $q$. The corollary is that in discs having $q\le\frac{2}{3}$, i.e. a shallow enough temperature profile, large grains in the Stokes regime at small Reynolds numbers remain in the disc. Such a criterion is applicable for real protoplanetary discs.

Second, if  $R_{\mathrm{g}} >800$, the drag force is quadratic in $\textbf{v} - \textbf{v}_{\mathrm{g}}$. Assuming that the radial motion is decoupled from the vertical motion, we perform the NSH expansion at small pressure gradient (cf. Eq. (\ref{dvptnaka})). We find that whatever the integer $j$, $\left(\etaz R^{-p} \right)^{j}  \left(\textbf{v} - \textbf{v}_{\mathrm{g}} \right) \to 0 $ at the limit $\etaz \to 0$. This means that both $v_{r}$ and $v_{\theta} - \sqrt{1/R}$ are flat functions as  their Taylor series expansion equals zero at each order. Consequently, they can not be determined by perturbation analysis. This property comes from the quadratic dependency of the drag with respect to the differential velocity and thus is not related to the grain size. Consequently, in this drag regime, the drag force is extremely efficient and the corrections to the Keplerian motion are negligible at every order of the perturbative expansion. The particles are very well coupled to the gas and do not migrate significantly.

Third, for the intermediate case, we could not manage to perform the expansion at small pressure gradients. However, we expect an intermediate behaviour between the two Stokes regime at small and large Reynolds numbers. Consequently, if $R_{\mathrm{g}} >1$, the migration motion becomes less efficient as the drag force is no longer linear with respect to the differential velocity between the gas and the dust particles. Thus, the main constraint for the radial-drift barrier due to the Stokes drag comes from the low Reynolds number regime for which the migration motion is the most efficient.

Finally, confusion often arises when defining the ``radial-drift barrier'' as Óthe difficulty a grain has of ``overcoming $s = s_\mathrm{opt}$'' (i.e.) reaching the B-mode. Indeed, as we have shown, grains can survive their migration motion in the Epstein regime when they are in the A-mode whenever $-p+q+1/2<0$, and grains can start their migration motion in the B-mode but be accreted in a finite time if $-p+q+1/2>0$. This study also shows for the Stokes regime that a grain ends its migration motion in the B-mode. However, as demonstrated in this work, the ability of the grain to overcome the radial drift barrier is only linked to the value of $q$. If $q>2/3$, the grain will be accreted onto the central star in a finite time, even if it has $s>s_\mathrm{opt}$. Thus, we would argue that the definition of the radial-drift barrier has to remain Òthe ability of the grain to be accreted onto the central star or depleted from the disc within its lifetimeÓ.

\section{Limitations of the model}
\label{Sec:Discuss}

We have demonstrated that the time it takes for grains to reach smaller and smaller radii increases dramatically under certain conditions. Specifically, if $-p+q+\frac{1}{2}<0$ (resp. $q<\frac{2}{3}$), grains experience a pile-up in the Epstein (resp. Stokes) regime. However, the model developed for the radial evolution of dust grains in this paper remains simple in that we neglect several important physical processes: turbulence, grain growth, collective motion of dust grains, dust feedback on the gas surface density and temperature profiles. We now discuss how those processes can modify the criteria derived above.
\begin{enumerate}

\item The local pressure maxima created by turbulence \citep{Cuzzi2001,Cuzzi2008} and the collective effects due to the dust drag onto the gas phase \citep{Youdin2005} are known to slow down the dust particles. However, the efficiency of these processes --- such as the non linearity of the streaming instability in global disc models and the life time of the pressure maxima --- in real discs remains difficult to quantify. Omitting these phenomena constitutes therefore an upper limit for the grain migration efficiency, which will be slowed by these additional processes.

\item In this study, we assume that changing the dust distribution does not change the thermal profile of the disc. We also neglect the viscous evolution of the disc, assuming that the viscous timescales are larger than the characteristic timescales of the initial dust evolution. It implies that we assume that $p$ is constant during the whole grains evolution. We can expect that for long term evolution, the surface density profile will flatten, leading to a smaller value of $p$. This makes the Epstein criterion harder to be met, while the Stokes criterion is not affected.

\item We have shown that even if the velocity of the grain's inward motion depends on their sizes, their outcome only depends on the surface density and temperature profiles. Thus, if we now consider growing (or fragmenting) grains, we expect that (\textit{i}) the intensity of the inward motion depends on the growth efficiency (this point will be discussed in detail in a forthcoming paper), but that (\textit{ii}) the grains outcome remains determined by our criteria, regardless of the growth regime.

\end{enumerate}

Following this discussion, our simple model deals with processes that are optimized for the grains to be depleted on the central object. Consequently, our Epstein and Stokes criteria for the radial-drift barrier constitute the least favourable limit for grain survival. Thus, we are confident when claiming that the radial-drift barrier does not occur in some classes of discs.  One may however expect that more discs are retaining their grains due to the complementary processes mentioned above.

It should be noted that the criterion $-p+q+1/2$ quantifies the outcome of the radial-drift motion of the grains, but not their kinematics (which depends on the grain size, the grain density, etc...). Thus, we provide predictions for which discs will retain the largest mass of solid particles, but do not predict for which discs the radial migration to the inner disc regions is the fastest. Full simulations like those developed in \citet{Brauer2008} are required to make predictions of the dust kinetics, even more so when the grain size evolution is driven by a complex model of growth and fragmentation. However, this study suggests that while complex simulations are useful to study the details of the dust dynamics, they are not required to determine the grains outcome.

As a conclusion of this section, we have mentioned that the physics treated by our model is not exhaustive. In real discs, the limits $-p+q+\frac{1}{2} = 0$ and $q = \frac{2}{3}$ may be softened by the effects of additional physical phenomena. However, these neglected processes (such as turbulence and grain growth) tend to decrease the efficiency of the dust radial motion. Our predictions of when the ``radial-draft barrier'' does not occur therefore remain valid. Our model represents a powerful indicator for predicting the dust behaviour in discs with given power law profiles: we expect that (i) discs satisfying $-p+q+\frac{1}{2}\le0$ retain their small grain population and that (ii) discs satisfying  $q\le\frac{2}{3}$ keep their large solids. On the contrary, discs for which $-p+q+\frac{1}{2}>0$ (resp.\ $q>\frac{2}{3}$) likely lose their small (resp. large) particles.

%
\section{Application to observed discs and planet formation}
\label{Sec:PlanetFormCsq}

\subsection{Validity of the criteria in real protoplanetary discs}
\label{Sec:RealDiscs}

We now study how the criteria we derived can be applied when considering the physical evolution of grains in observed discs, with finite inner radii and finite lifetimes. The analytic expressions of the previous sections have been derived using dimensionless quantities. We now provide the physical timescales of the radial dust motion estimating the parameters involved in real protoplanetary discs. We consider in this section a typical CTTS disc, of mass $M_\mathrm{disc}=10^{-2}\ M_\odot$ around a 1~$M_\odot$ star, extending from $r_\mathrm{in}=10^{-2}$~AU to $r_\mathrm{out}=10^3$~AU. The disc inner edge is chosen to correspond to the dust sublimation radius for a 1~$M_\odot$ star, whereas its outer boundary is representative of the largest observed discs. Its vertical extent is set by the choice of the temperature scale. We take $T(1$~AU$)=150$~K, a typical value obtained by \citet{AWOph2007} in their disc observations.

\begin{figure}
\begin{center}
\resizebox{\hsize}{!}{\includegraphics[angle=270]{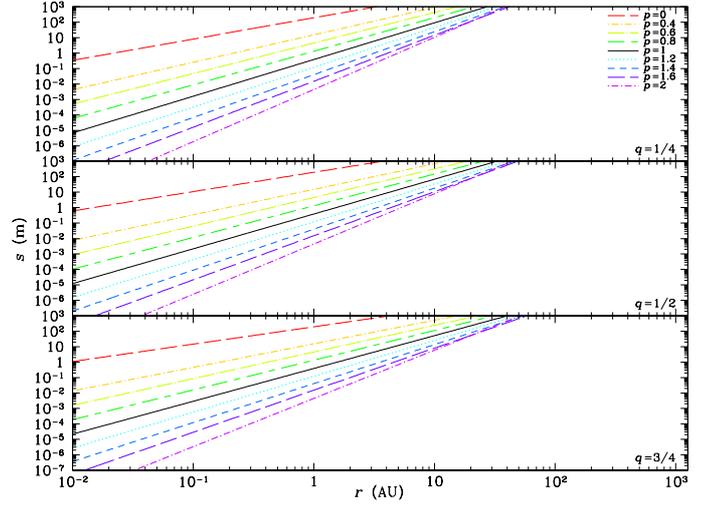}}
\caption{Transition between the Epstein and the Stokes regimes in a protoplanetary disc of $M_{\rm{disc}} = 10^{-2} M_\odot$ extending from $r_{\rm{in}} = 10^{-2}$ AU to $r_{\rm{out}} = 10^{3}$ AU for several values of $p$ and $q$. Grains with sizes below (resp. above) than the curve experience the Epstein (resp. Stokes) drag regime.}
\label{FigTrans}
\end{center}
\end{figure}

The transition from the Epstein to the Stokes regime occurs when $\lambda_{\mathrm{g}} = \frac{4s}{9}$, or
\begin{equation}
s=\frac{9}{4}\lambda_\mathrm{g}= \frac{45\pi^\frac{3}{2}}{256}\frac{m H'_0}{\sigma_\mathrm{s}\Sigma'_0}\,r^{p-\frac{q}{2}+\frac{3}{2}},
\end{equation}
and is represented in the $(r,s)$ plane in Fig.~\ref{FigTrans} for this typical disc for different values of the surface density and temperature power-law exponents $p$ and $q$. The Stokes regime is seen to apply to large bodies in the disc inner regions and for large values of both $p$ and $q$.

\begin{figure}
\begin{center}
\resizebox{\hsize}{!}{\includegraphics[angle=-90]{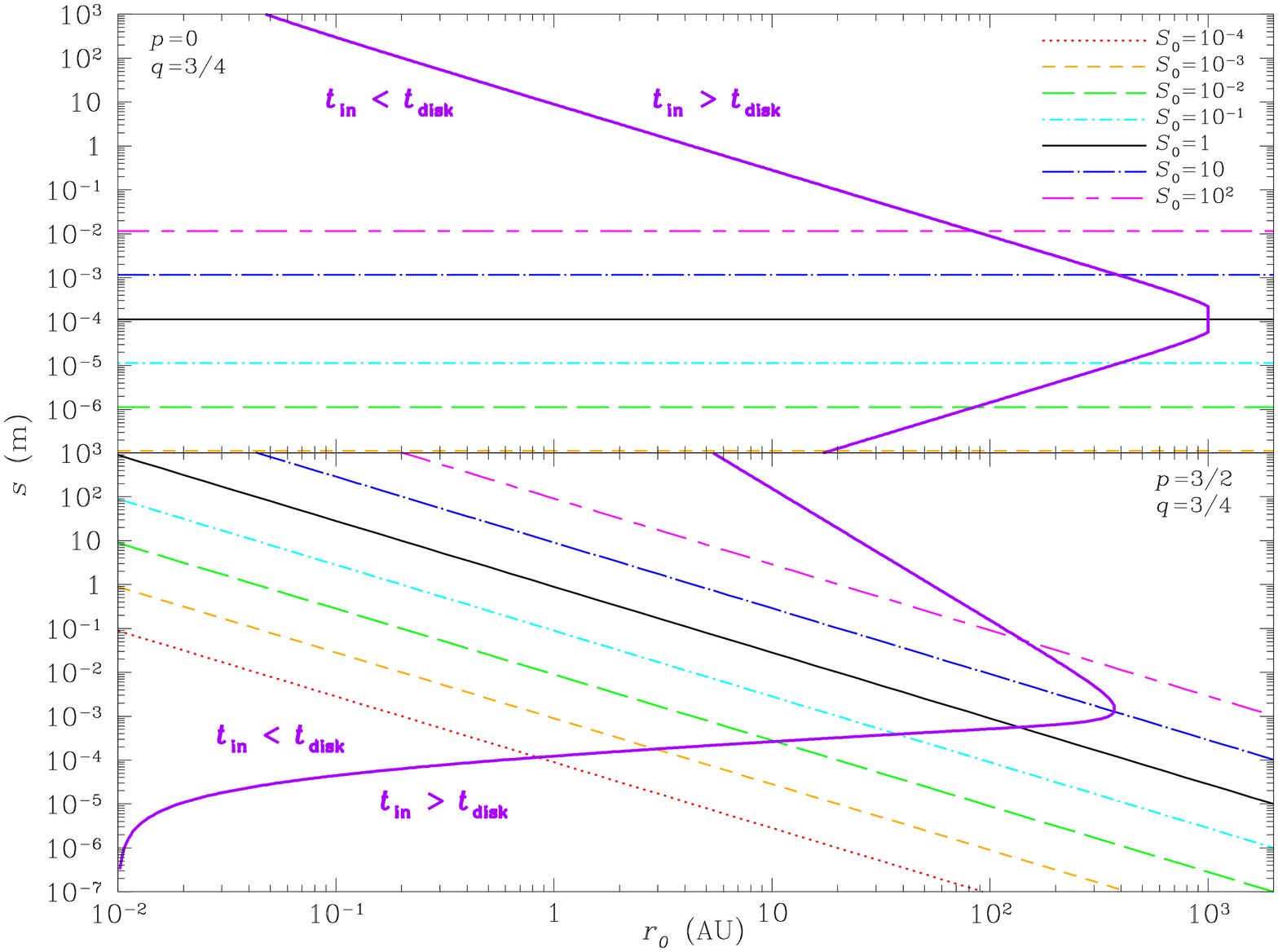}}
\caption{Values of $\sz$ as a function of grain size $s$ and initial position $r_0$ for a disc of mass $M_\mathrm{disc}=0.01\ M_\odot$ extending from $r_\mathrm{in}=10^{-2}$~AU to $r_\mathrm{out}=10^{3}$~AU, with grains of intrinsic density $\rho_\mathrm{d}=1000$~kg\,m$^{-3}$ and $\etaz=10^{-2}$, for $p=0$ and $q=\frac{3}{4}$ (top) and $p=\frac{3}{2}$ and $q=\frac{3}{4}$ (bottom). The thick line shows the limit between grains that are accreted onto the star ($t_\mathrm{in}<t_\mathrm{disc}$) and those that survive in the disc ($t_\mathrm{in}>t_\mathrm{disc}$), i.e. the survival limit, in the Epstein regime.}
\label{FigLimTinTdisc}
\end{center}
\end{figure}

Disc lifetimes are generally thought to be a few Myr \citep{Haisch2001,Carpenter2005}, and thus we take $t_\mathrm{disc}\sim10^6$~yr. For a grain starting at a distance $r_0$ from a 1~$M_\odot$ star, the dimensionless value $T_\mathrm{disc}$ is therefore
\begin{equation}
T_\mathrm{disc}=\frac{t_\mathrm{disc}}{\tkz}
=\frac{\sqrt{\gm_\odot}\,t_\mathrm{disc}}{r_0^{3/2}}
\sim\frac{6\times10^6}{[r_0\ \mathrm{(AU)}]^{3/2}}.
\label{EqTdisc}
\end{equation}
The dimensionless value of the dust disc inner radius ($r_\mathrm{in}\sim0.01$~AU) for a grain starting at $r_0$ is
\begin{equation}
R_\mathrm{in}=\frac{r_\mathrm{in}}{r_0}\sim\frac{10^{-2}}{r_0\ \mathrm{(AU)}}.
\label{EqRin}
\end{equation}
The link between dimensionless and real grain sizes is made through the optimal size for radial migration. Considering first the Epstein regime, its midplane value has a radial dependence given by
\begin{equation}
s_\mathrm{opt}(r,0)=\frac{\Sigma(r)}{\sqrt{2\pi}\,\rho_\mathrm{d}}=
\frac{\Sigma'_0\,r^{-p}}{\sqrt{2\pi}\,\rho_\mathrm{d}}.
\label{EqSopt}
\end{equation}
The dimensionless size of a particle of size $s$ starting its migration at position $r_0$ is therefore
\begin{equation}
\sz=\frac{s}{s_{\mathrm{opt},0}}=\frac{\sqrt{2\pi}\,\rho_\mathrm{d}}
{\Sigma'_0}\,s\,r_0^p.
\label{EqSSopt}
\end{equation}
Values of $\sz$ are plotted in the $(r_0,s)$ plane in Fig.~\ref{FigLimTinTdisc} for a disc of total mass $M_\mathrm{disc}=0.01\ M_\odot$ extending from $r_\mathrm{in}=10^{-2}$~AU to $r_\mathrm{out}=10^{3}$~AU, with grains of intrinsic density $\rho_\mathrm{d}=1000$~kg\,m$^{-3}$ and $\etaz=10^{-2}$, for both $p=0$ and $p=\frac{3}{2}$.

The dimensionless time $T_\mathrm{in}$ for a grain to reach $R_\mathrm{in}$ is given by Eq.~(\ref{implicitNSH}). In combination with Eq.~(\ref{EqRin}), this gives an expression of $T_\mathrm{in}$ as a function of $\sz$ and $r_0$, whereas Eq.~(\ref{EqTdisc}) gives an expression of $T_\mathrm{disc}$ as a function of $r_0$. Equating them yields a second order equation in $\sz$ as a function of $r_0$, which can be solved to determine under which conditions a grain reaches the disc inner edge at the end of its lifetime. Using Eq.~(\ref{EqSSopt}) gives the corresponding relationship between the grain size and its initial position:
\begin{equation}
\begin{array}{l}
s=\dst\frac{p+q+\frac{1}{2}}{2\sqrt{2\pi}\,\rho_\mathrm{d}}
\frac{\Sigma'_0\,r_0^{-p}}{1-\left(\frac{r_\mathrm{in}}{r_0}\right)^{p+q+\frac{1}{2}}}
\left[\frac{\sqrt{\gm}\,t_\mathrm{disc}\,\etaz}{r_0^{3/2}}\right. \\[5ex]
\dst\left.\pm\sqrt{\frac{\gm\,t_\mathrm{disc}^2\,\etaz^2}{r_0^3}
-4\frac{\left(1-\left(\frac{r_\mathrm{in}}{r_0}\right)^{-p+q+\frac{1}{2}}\right)\left(1-\left(\frac{r_\mathrm{in}}{r_0}\right)^{p+q+\frac{1}{2}}\right)}
{\left(-p+q+\frac{1}{2}\right)\left(p+q+\frac{1}{2}\right)}}\right],
\end{array}
\label{EqLimTinTdisc}
\end{equation}
which is plotted as a thick line in Fig.~\ref{FigLimTinTdisc}. It separates the $(r_0,s)$ plane into regions in which grains reach $r_{\mathrm{in}}$ and leave the disc before it dissipates ($t_\mathrm{in}<t_\mathrm{disc}$) or survive in the disc throughout its lifetime ($t_\mathrm{in}>t_\mathrm{disc}$). We call this curve the survival limit.

For $-p+q+\frac{1}{2} > 0$, illustrated by the case $(p=0,q=\frac{3}{4})$, Fig.~\ref{FigLimTinTdisc} shows as expected that most of the grains are lost during the disc lifetime. However, small and large grains initially in the outer disc survive, therefore even with this profile, the disc retains a fraction of its grain population before it dissipates. Moreover, one may expect growing grains that reach $S_0=1$ to be inevitably accreted onto the star (unless the growth process is fast enough for grains to outgrow the fast migrating sizes before they leave the disc, see \citealt{Laibe2008}). Such discs may not form planets, but their remaining dust content may still make them observable, although they would likely be faint. 

For $-p+q+\frac{1}{2} \le 0$, illustrated by the case $(p=\frac{3}{2}, q=\frac{3}{4})$, even though all grains fall on the star in an infinite time, some of them reach the disc's inner edge before it dissipates. On the contrary, for $r_0>350$~AU, grains of all sizes remain in the disc. This is also the case for all (sub)micron-sized grains, whatever their initial location, as well as most of the grains up to 0.1~mm. These grains likely make the disc bright and easy to observe, since they are the grains contributing most to the disc emission at IR and submm wavelengths. A large reservoir of grains is available to participate in the planet formation process, however a firm conclusion on their survival as their size evolves would require incorporating a treatment of grain growth, as discussed in Sect.~\ref{Sec:Discuss}.

It should be noted that the disc used in these examples represents a lower limit, as it is low mass and very extended. A more massive disc with a smaller outer radius would have a larger surface density, and the corresponding survival limit in Fig.~\ref{FigLimTinTdisc} would be shifted vertically towards larger sizes and more and more grains of larger sizes would survive.

\begin{figure}
\begin{center}
\resizebox{\hsize}{!}{\includegraphics[angle=0]{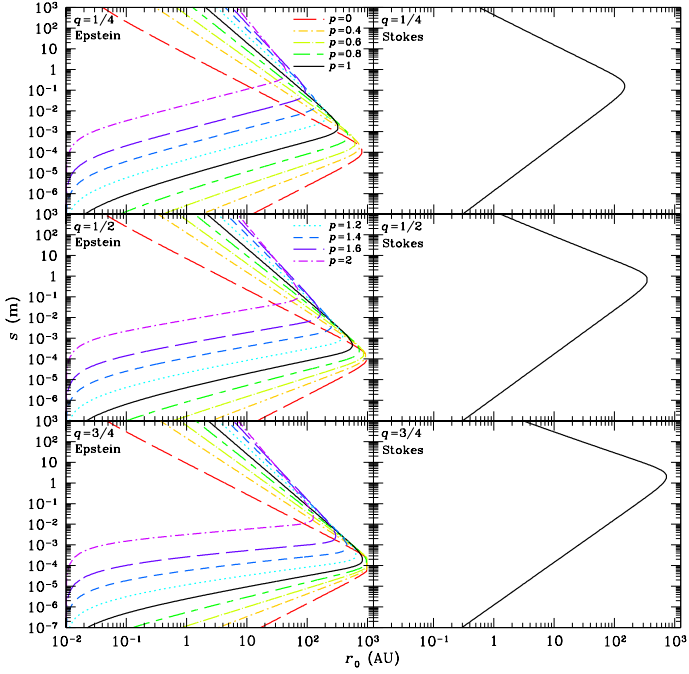}}
\caption{Survival limits of grains for different values of $p$ and $q$. Left: Epstein regime, right: Stokes regime. Grains to the left of the curves ($t_\mathrm{in}<t_\mathrm{disc}$) are accreted onto the star whereas those to the right ($t_\mathrm{in}>t_\mathrm{disc}$) survive in the disc.}
\label{FigLim2TinTdisc}
\end{center}
\end{figure}

The left panels of Fig.~\ref{FigLim2TinTdisc} show the influence of the surface density and temperature profiles on the location and shape of the survival limit curve in the $(r_0,s)$ plane in the Epstein regime. Increasing $p$ from 0 to 2 shifts the curve towards smaller radii and larger grain sizes, as well as slightly tilts it clockwise. The outer disc region in which grains of all sizes survive extends inwards, as well as the surviving population of small grains as the curve's lower branch shifts upwards and becomes flatter. On the contrary, the steepening of the curve's upper branch, confining the population of surviving large solids to the disc outer regions, is less dramatic.
Increasing $q$ from $\frac{1}{4}$ to $\frac{3}{4}$ also tilts the curve clockwards, but shifts it towards larger radii and smaller grain sizes. However, its effect is more limited than that of changing $p$.
A disc with a steeper surface density profile and a shallower temperature profile is therefore more efficient at retaining a larger quantity of small grains and up to larger sizes. Indeed, large $p$ and small $q$ values are required to meet the $-p+q+\frac{1}{2}<0$ criterion introduced in Sect.~\ref{Sec:RadialMotion}.

An equation very similar to Eq.~(\ref{EqLimTinTdisc}) can be obtained for the linear Stokes regime by replacing $p$ and $q$ by $p'=\frac{q-3}{2}$ and $q'=q$ (since the equation of motion has the same structure for both drag regimes, see Sect.~\ref{Sec:Stokes}), and using the expression of $s_{\mathrm{opt},0}$ for that regime, given in Table~\ref{tabdrag}. Here $s_{\mathrm{opt},0}\propto T^\frac{1}{4}$: the weak temperature dependence results in very little change for a large range of temperatures below or above our adopted value of $T(1$~AU$)=150$~K. The right panels of Fig.~\ref{FigLim2TinTdisc} show the survival limit in the Stokes regime for the different values of $q$ (note that it no longer depends on $p$). When $q$ increases, the curve's lower branch slides towards larger radii, making the survival of particles in the Stokes regime less and less favourable.

Given the form of Eq.~(\ref{EqLimTinTdisc}) and its different expressions for each drag regime, it is not possible to compute analytically the survival limit for a grain transitioning from the Epstein to the Stokes regime as it migrates inwards. However, large values of $p$ and $q$, for which the Stokes region is the largest, are not observed in protoplanetary discs (see Sect.~\ref{Sec:Obs}), and in practical cases the Stokes regime only applies to a small area of the $(r_0,s)$ plane. Small grains, which are detected in disc observations at IR and sub-millimetre (submm) wavelengths, are mostly subject to the Epstein drag. We therefore focus on that regime in the following.

\begin{figure}
\begin{center}
\resizebox{\hsize}{!}{\includegraphics{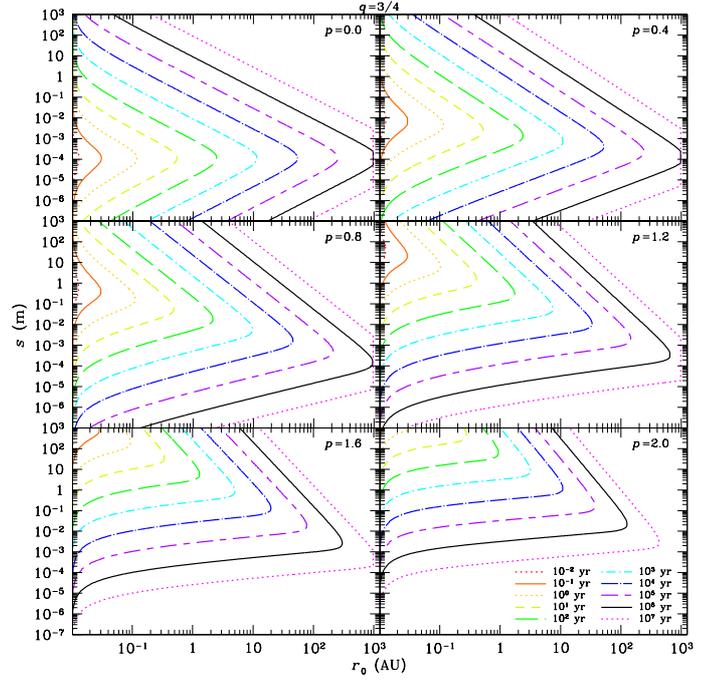}}
\caption{Isocontours of the survival time (i.e. the time needed to reach the disc inner edge at $r_\mathrm{in}=0.01$~AU) of grains of size $s$ and initial position $r_0$ for $q=\frac{3}{4}$ and different values of $p$.}
\label{FigLimTdisc}
\end{center}
\end{figure}

Equation~(\ref{EqLimTinTdisc}) can give quantitative information about the outcome of the grain population. Replacing $t_\mathrm{disc}$ by any time $t$ gives the location in the $(r_0,s)$ plane of grains reaching the disc inner edge (at $r=r_\mathrm{in}$) in that time $t$, which is therefore the survival time of those grains. Its isocontours are shown in Fig.~\ref{FigLimTdisc} for different values of $p$ and for $q=\frac{3}{4}$. Only one value of $q$ is shown as the $q$ dependence is moderate, as can be seen from the left panels of Fig.~\ref{FigLim2TinTdisc}. The fate of particular dust grains can easily be obtained from these figures. For example, in the context of disc observations, 1~mm grains initially at 100~AU fall on the $10^5$~yr contour for $p=0$. Their survival time decreases to a few $10^4$~yr for $p\sim0.8$, and increases again to values larger than $10^6$~yr as $p$ increases. At an initial position of a few hundred AU, 1~mm grains survive longer than $10^6$~yr for any $p$, therefore long enough to contribute to the disc emission over its entire lifetime. As noted above, such grains have longer survival times for higher $p$ and lower $q$ values. As another example, in the context of planetesimal formation, the survival time of a 1~m particle initially at 1~AU is $\sim10^5$~yr for $p=0$, decreases to $\sim10^2$~yr for $p\sim1$, and increases again to $\sim10^3-10^4$~yr for $p=2$. The ability of such particles to remain for long enough in the disc to grow to larger sizes therefore strongly depends on the surface density profile. As a general rule, the survival of pre-planetesimals in the inner disc is favoured by small values of $p$.

\begin{figure}
\begin{center}
\resizebox{\hsize}{!}{\includegraphics{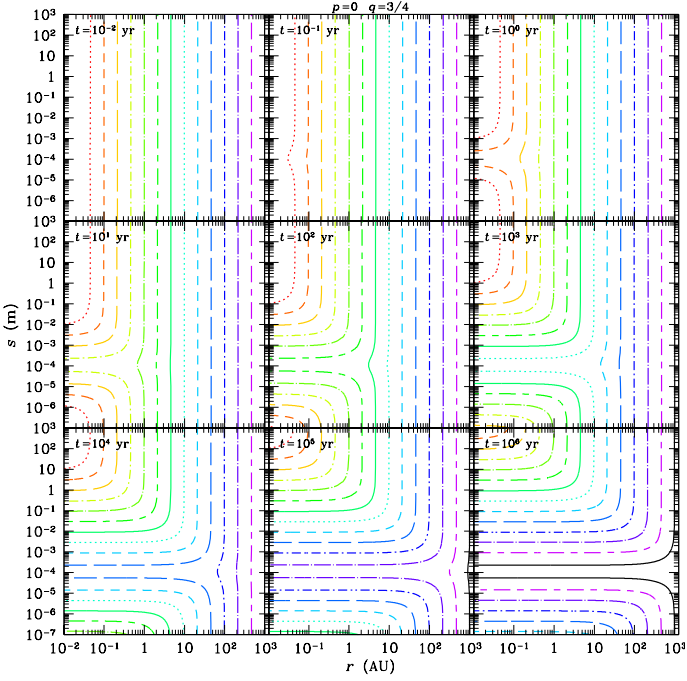}}
\caption{Time evolution (in nine snapshots from $t = 10^{-2}$ to $10^{6}$ yr) of isocontours of the initial position of grains in the $(r,s)$ plane for the disc with $p=0$ and $q=\frac{3}{4}$. The label for each contour can be deduced from its abscissa in the upper left panel at $t = 10^{-2}$ yr.}
\label{FigLimEvol_p0}
\end{center}
\end{figure}

\begin{figure}
\begin{center}
\resizebox{\hsize}{!}{\includegraphics{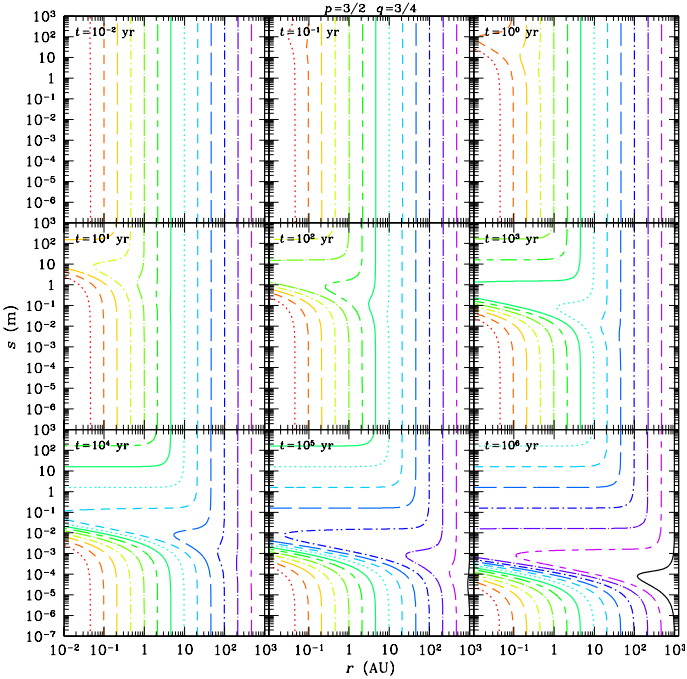}}
\caption{Same as Fig.~\ref{FigLimEvol_p0} for $p=\frac{3}{2}$ and $q=\frac{3}{4}$.}
\label{FigLimEvol_p1.5}
\end{center}
\end{figure}

Similarly, replacing now $r_\mathrm{in}$ with any radius $r$ in Eq.~(\ref{EqLimTinTdisc}) gives the locus in the $(r_0,s)$ plane of grains reaching that radius $r$ at any time $t$. Alternatively, one can plot isocontours of the initial position $r_0$ of grains in the $(r,s)$ plane at various times, thus showing the radial evolution of grains with the same initial position but different sizes. This is shown in Fig.~\ref{FigLimEvol_p0} for ($p=0$, $q=\frac{3}{4}$) and Fig.~\ref{FigLimEvol_p1.5} for ($p=\frac{3}{2}$, $q=\frac{3}{4}$). These plots make it easy to compare the radial evolution of any particle to any physical timescale of interest in the disc. In particular, they show that the disc still contains particles at all radii at the end of its evolution ($t=10^6$~yr). No grains are found to the right of the $r_0=10^3$~AU contour, since this was the initial outer disc radius. In the disc with $(p=0,q=\frac{3}{4})$, no grains between $\sim0.06$ and $\sim0.2$~mm remain, and grains of other sizes still present were initially in the outer disc. Given that the grains of sizes which contribute to IR and submm emission have come from a small fraction of the initial disc, this disc is likely faint. In the disc with $(p=\frac{3}{2},q=\frac{3}{4})$, only grains with $s\sim0.1$~mm are absent from the very outer regions, and the observable grains come from a larger portion of the disc, likely making the disc brighter than in the previous case.

As a conclusion, the analytic criteria derived above apply even when taking into account the finite lifetime (or inner radius) of the disc. For most CTTS discs, the dust is in the Epstein drag regime (except for some extreme values for the grains sizes and discs profiles). Therefore, the grain's radial outcome is given by the value of $-p+q+\frac{1}{2}$. However, the transition between discs for which the radial-drift barrier occurs or not consists more of a continuum around the value $-p+q+\frac{1}{2}=0$ than in the sharp transition predicted by the analytic model. Therefore, the radial motion of the grains has to be studied on a case-by-case basis for discs close to the transition $-p+q+\frac{1}{2} = 0$, using the figures shown above in this section.

\subsection{Constraining physical systems}
\label{Sec:Obs}

\begin{figure}
\resizebox{\hsize}{!}{\includegraphics[angle=-90]{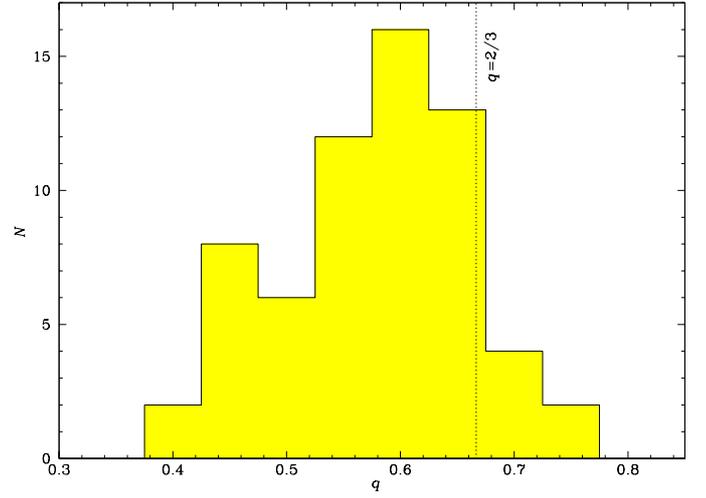}}
\caption{Histogramm of the $q$ parameters obtained from \citet{AWTau2005,AWOph2007} data of 63 observed discs. The distribution is roughly comprised between 0.4 and 0.8 and, centred around 0.55. Approximately 90 \% of the discs satisfy $q<2/3$.}
\label{histodiscs}
\end{figure}

\begin{figure}
\resizebox{\hsize}{!}{\includegraphics{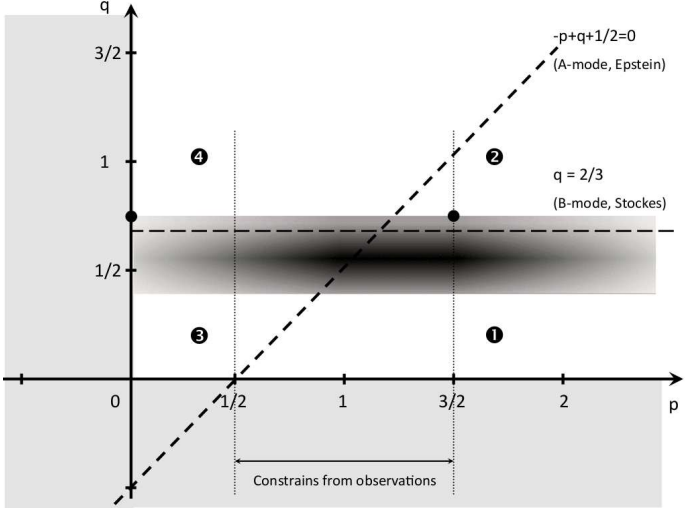}}
\caption{Location of the different outcomes of radial migration in the ($p$,$q$) plane. Dashed (resp. dotted) line: limit between accretion without or with grains pile-up resulting in a finite or infinite time in the A-mode of the Epstein regime (resp. B-mode of the Stokes regime at small Reynolds numbers). Shaded area: location of observed discs. Black dots: discs used as examples in Sect.~\ref{Sec:Asymptotic}.}
\label{pqdistrib}
\end{figure}

We now turn to observed discs and check if they meet our Epstein and Stokes criteria to determine whether the radial-drift barrier is constraining for planet formation. To estimate the values of the $p$ and $q$ exponents for real discs, we use the results of disc modeling obtained by \citet{AWTau2005,AWOph2007} from data on 63 discs in $\rho$ Ophiuchi, Taurus and Aurigae. Using sub-millimetre fluxes measured at several wavelengths, they fit a range of disc parameters assuming a geometrically thin irradiated disc with opacities from \citet{Beckwith1990}, a gas-to-dust ratio of 100, a disc radius of 100 AU and zero disc inclination. The temperature exponent $q$ is well constrained by the observational data set: the histogram of most probable $q$ values is shown in Fig.~\ref{histodiscs}.
However, $p$ is not well constrained and is usually assumed to be $\frac{3}{2}$. Very flat profiles with $p<\frac{1}{2}$ and very steep profiles with $p>\frac{3}{2}$ seem to be excluded \citep{Dutrey1996,Wilner2000,Kitamura2002,Testi2003,Isella2009,AWOph2007,Andrews2009}.  We represent the disc distribution modeled by Andrews \& Williams from observations in the $\left(p,q\right)$ diagram of Fig.~\ref{pqdistrib}: the histogram of Fig.~\ref{histodiscs} is represented by the gray-shaded area and spread over a range of $p$ values, taking into account that extreme values of $p$ are less probable. The dashed line ($-p+q+1/2=0$) represents the border between migration in an infinite time and accretion onto the central star for the A-mode of migration in the Epstein regime, while the thick dotted line ($q = 3/2$) represents that same border for the B-mode of migration in the Stokes regime at low Reynolds number. The two black circles indicate the discs used as examples in Sect.~\ref{Sec:Asymptotic}.
We have split the disc distribution in four regions in the $\left(p,q \right)$ plane:
\begin{enumerate}

\item region 1: $-p + q +\frac{1}{2} \le 0$ and $q \le \frac{2}{3}$: both small and large grains experience the pile-up effect. Those discs are potentially observable and may favour planet formation.
\item region 2: $-p + q +\frac{1}{2} \le 0$ and $q > \frac{2}{3}$: only small grains experience the pile-up effect: even though such discs retain their small grains, the population of pre-planetesimals in the disc inner regions may efficiently be accreted onto the central star (at least until they reach the high-$R_\mathrm{g}$ Stokes regime).

\item region 3: $-p + q +\frac{1}{2} > 0$ and $q \le \frac{2}{3}$: if the pre-planetesimals can form before the entire distribution of  small grains has been accreted onto the central object, they will remain in the disc and may constitute planet embryos.

\item region 4: $-p + q +\frac{1}{2} > 0$ and $q > \frac{2}{3}$: both small and large grains are accreted onto the central star.
\end{enumerate}

The Epstein criterion indicates that for $q$ values in the range constrained by observations, discs which keep their small grain population, and are therefore likely to be bright in the IR and submm, should have $p$ values approximately in the $[1;\frac{3}{2}]$ range. This is indeed what is found in most disc surveys \citep{Ricci2010a,Ricci2010b}. On the contrary, smaller $p$ values should correspond to discs which lose most of their small grains, and are therefore more difficult to detect. This is what is found by \citet{Andrews2010}, who pushed their previous observations of the Ophiuchus star forming region \citep{Andrews2009} down to fainter discs, finding for this new sample a median $p$ value of 0.9, lower than for brighter discs. The criterion we derive in this paper for small grains in the Epstein regime provides therefore the correct behaviour for explaining the range of $p$ values of observed discs. However, this result has to be considered carefully for two reasons. Firstly, the $p$ and the $q$ exponents determined from the observations have to be considered with their respective errors. Given these uncertainties, one may not be able to distinguish between a strict negative or positive value for $-p+q+\frac{1}{2}$. Second, the boundary between the different zones of the $(p,q)$ plane consists more of a continuum rather than a strict limit due to the finite lifetime/inner radii of the discs. The outcome of the grains may thus not be predicted when the value of $-p+q+\frac{1}{2}$ is close to zero.

Now turning to the Stokes criterion for large solids, Figs.~\ref{histodiscs} and \ref{pqdistrib} show that the vast majority of observed protoplanetary discs have shallow temperature profiles ($q\le\frac{2}{3}$) and are thus able to retain their population of pre-planetesimals. These discs are therefore relevant places to find evidence of planet formation, provided small grains can efficiently grow to form pre-planetesimals. For the remainder of the disc population, the outcome of pre-planetesimals will likely depend on their ability to reach the high Reynolds number Stokes regime. However, the case of a steep radial temperature profile can be encountered in at least one particular situation: circumplanetary discs which typically have temperature profiles with $q = 1$ \citep{Ayliffe2009}. In this environment, we predict from our Stokes criterion that planetesimals will be accreted onto the planet. The timescale of the planet formation by the core-accretion process, which usually corresponds to the time required to release the gravitational energy of the accreted bodies \citep{Pollack1996}, is thus increased as the drag from the gas onto the planetesimals releases an additional thermal contribution.

\section{Conclusion and perspectives}
\label{Sec:Conclusion}

In this study, we have generalised the radial grain motion studies of \citetalias{Weidendust1977}, \citetalias{Nakagawa1986} and \citetalias{YS2002} for both the Epstein and the Stokes regimes, taking into account the effects of both the surface density and temperature profiles in the disc. As observations do not provide direct information about the three dimensional structure of discs, radial profiles of surface density and temperature are often described by power laws: $\Sigma\left(r\right) = \Sigmap r^{-p}$ and $\mathcal{T}\left(r\right) = \mathcal{T}_{0}' r^{-q}$, where both $p$ and $q$ take positive values. The radial dust behaviour in those discs is governed by the competition between gravity and gas drag. The final outcome of the radial motion is set by two counterbalancing effects. First, the temperature increases when the radius decreases. Consequently, the deviation from the Keplerian velocity increases, which accelerates the dust's radial inward motion. At the same time, the surface density also increases, which increases the gas drag efficiency and slows down the dust motion. The competition between these two effects fixes the ultimate mode of migration of the grain (A-mode, where the drag dominates or B-mode, where the gravity dominates) and thus the final outcome for the dust motion. In this work, we have shown that it can be represented by an analytical criterion which depends on the drag regime. For the Epstein drag regime (in which the ultimate radial motion is in the A-mode), if $-p+q+\frac{1}{2} > 0$, the dust particle is accreted onto the central star in a finite time, and if $-p+q+\frac{1}{2} \le 0$, the grain pile-up results in an infinite accretion time and small dust grains remain in the disc. We have shown that, as expected, these conclusions are somewhat mitigated when taking into account the finite disc lifetime and finite disc size. However, the outcomes still remain similar: the bulk of the small grain population is lost to the star in the first case, whereas in the second case the disc keeps most of its small grains. A similar criterion is found for the Stokes regime at low Reynolds number: if $q \le \frac{2}{3}$, the accretion time is infinite and large pre-planetesimals remain in the disc and can constitute the primary material for planet formation. However, the Stokes radial motion differs from the Epstein regime as the ultimate radial motion occurs in the B-mode.

The observational consequence is that discs with a large population of small grains should be strong emitters in the infrared and sub-millimetre and should be easier to observe, and that those having lost most of their small dust should be fainter and harder to detect. This is indeed what is found: a large fraction of the observed discs have large $p$ values whereas fainter discs tend to have lower $p$ values \citep{Andrews2010}, in agreement with this Epstein criterion. In addition, most of the observed discs have $q \le \frac{2}{3}$, allowing them to retain also their large pre-planetesimals. As noted by \citet{Ricci2010a,Ricci2010b}, explaining the data requires a mechanism halting or slowing down the radial migration of dust grains. We show here that local pressure maxima need not be invoked, but rather that the combination of adequate surface density and temperature profiles is sufficient. The $p$ and $q$ exponents used to reach our conclusions are of course strongly dependent on the model used to fit the data. However, even varying the fitting models, a large majority of discs still satisfy both conditions $-p+q+\frac{1}{2} \le 0$ and $q \le \frac{2}{3}$. Consequently, the radial-drift barrier (or the so-called metre-size barrier when considering an MMSN disc) does not appear to constitute a problem for planet formation for the discs that we do observe.

Our conclusions presented in this study assumed that the grain size remains constant during its motion. However, observations tell us that grains do grow \citep{Testi2003,Wilner2003,Apai2005,Lommen2007,Lommen2009}. Grain growth is studied in various theoretical studies \citep{Schmitt1997, StepVal1997, Suttner1999, Tanaka2005, DullemondDom2005, Klahr2006, Garaud2007, Brauer2008, Laibe2008, Birnstiel2009}. In a forthcoming paper, we will generalise the formalism developed here to explain the radial and vertical behaviour of growing dust grains.

\begin{acknowledgements}
This research was partially supported by the Programme National de Physique
Stellaire and the Programme National de Plan\'etologie of CNRS/INSU, France,
and the Agence Nationale de la Recherche (ANR) of France through contract
ANR-07-BLAN-0221. The authors want to thank C.~Terquem, L.~Fouchet, S.~Arena and E.~Crespe for useful comments and discussions. We also thank the referee for greatly improving the quality of this work by suggesting we include the important Stokes regime and present real numbers (time scales and grain sizes) for our criteria.
\end{acknowledgements}

\bibliographystyle{aa}
\bibliography{bibliodelta}

\begin{appendix}
\section{Notations}
\label{App:Notations}

The notations and conventions used throughout this paper are summarized in Table~\ref{tabnote}.
\begin{table}
\begin{center}
\begin{tabular}{ll}
\hline Symbol & Meaning \\ \hline
$M$ & Mass of the central star \\
$\textbf{g}$ & Gravity field of the central star \\ 
$\Rz$ & Initial distance to the central star \\
$\rhog$ & Gas density \\
$\brhog\left(r\right)$ & $\rhog\left(r,z=0\right)$ \\
$\cs$ & Gas sound speed \\
$\bcs\left(r\right)$ & $\cs\left(r,z=0\right)$ \\
$\csz$ & Gas sound speed at $\Rz$ \\
$T$ & Dimensionless time\\
$\mathcal{T}$ & Gas temperature ($\mathcal{T}_{0}$: value at $\Rz$)\\ 
$\Sigmaz$ & Gas surface density at $\Rz$ \\
$p$ & Radial surface density exponent \\
$q$ & Radial temperature exponent \\
$P$ & Gas pressure \\
$v_{\mathrm{k}}$ & Keplerian velocity at $r$ \\
$\vkz$ & Keplerian velocity at $\Rz$ \\
$\Hz$ & Gas scale height at $\Rz$ \\
$\phiz$ & Square of the aspect ratio $\Hz/\Rz$ at $\Rz$ \\
$\etaz$ & Sub-Keplerian parameter at $\Rz$ \\
$s$ & Grain size \\
$S$ & Dimensionless grain size \\
$\sz$ & Initial dimensionless grain size \\
$y$ & Grain size exponent in the drag force  \\
$\textbf{v}_{\mathrm{g}}$ & Gas velocity \\
$\textbf{v}$ & Grain velocity \\
$\rhod$ & Dust intrinsic density \\
$\md$ & Mass of a dust grain \\
$\ts$ & Drag stopping time \\
$\tsz$ & Drag stopping time at $\Rz$ \\
\hline
\end{tabular}
\end{center}
\caption{Notations used in the article.}
\label{tabnote}
\end{table}
%
\section{disc structure}
\label{App:discStructure}

\subsection{Hydrostatic equilibrium}

At stationary equilibrium ($\frac{\partial}{\partial t} = 0$), gas velocities $\textbf{v}_{\mathrm{g}r}$, $\textbf{v}_{\mathrm{g}\theta}$, $\textbf{v}_{\mathrm{g}z}$ and the gas density $\rhog$ obey mass conservation and the Euler equation:
\begin{equation}
\left\lbrace 
\begin{array}{rcl}
\dst \frac{1}{r} \frac{\rhog \partial r v_{\mathrm{g}r}}{\partial r} +  \frac{\rhog \partial v_{\mathrm{g}z}}{\partial z} & = & \dst 0 ,\\
\dst \rhog \textbf{v}_{\mathrm{g}}.\nabla \textbf{v}_{\mathrm{g}} & = & \dst -\nabla P + \rhog \textbf{g} .
\end{array}
\right.
\label{startequgas}
\end{equation}
The solution of Eq.~(\ref{startequgas}) requires:
\begin{equation}
v_{\mathrm{g}r} = v_{\mathrm{g}z} =0 ,
\label{vgasnul}
\end{equation}
which ensures mass conservation. Projecting the Euler equation on $\textbf{e}_{z}$:
\begin{equation}
\frac{1}{\rhog}\frac{\partial P}{\partial z} = -\frac{\gm z}{\left( \dd \right)^{3/2} } .
\label{eqhydro}
\end{equation}
Assuming that:
\begin{equation}
P = \cs^{2}\left(r,z\right) \rhog ,
\label{sspeed}
\end{equation}
and dividing both sides of Eq.~(\ref{eqhydro}) by $\cs^{2}$, we have:
\begin{equation}
\frac{\partial \, \mathrm{ln}\left(\cs^{2} \rhog \right)}{\partial z} = - \frac{\gm z}{\left(\dd \right)^{3/2} \cs^2}.
\label{eqhydroext}
\end{equation}
Integrating Eq.~(\ref{eqhydroext}) between 0 and $z$ provides:
\begin{equation}
\rhog \left(r,z\right) = \frac{\bp\left(r\right)}{\cs^2\left(r,z\right)}\mathrm{e}^{\displaystyle -\int_{0}^{z}\frac{\gm z' \mathrm{d}z'}{\left(r^2 + z'^2 \right)^{3/2} \cs^{2}\left(r,z'\right) }}.
\label{eqhydroint}
\end{equation}
This expression can be simplified by the following approximations:
\begin{itemize}
\item In the special vertically isothermal case, where the sound speed depends only on the radial coordinate, Eq.~(\ref{eqhydroint}) simplifies to:
\begin{equation}
\rhog \left(r,z\right) = \brhog \left(r\right) \mathrm{e}^{\displaystyle - \frac{\gm}{\bcs^{2}\left(r\right)} \left[\frac{1}{r} - \frac{1}{\sqrt{r^{2}+z^{2}}} \right]} .
\label{simpcsr}
\end{equation}
\item Further, assuming a thin disc ($\left(\frac{z}{r} \right)^{2}\ll 1$), a Taylor series expansion of Eq.~(\ref{simpcsr}) leads to:
\begin{equation}
\rhog \left(r,z\right)= \brhog \left(r\right) \mathrm{e}^{-\frac{z^{2}}{2H\left( r \right)^{2}}}, 
\label{simpangles}
\end{equation}
with:
\begin{equation}
H\left( r \right) = \frac{r \bcs \left(r\right)}{\vk \left( r \right)}, 
\label{Hscale}
\end{equation}
which is the classical scale height for vertically isothermal thin discs.
\end{itemize}

\subsection{Azimuthal velocity}
The radial component of the Euler equations is given by:
\begin{equation}
-\frac{\vgtheta^{2}}{r} = -\frac{1}{\rhog} \frac{\partial P}{\partial r} - \frac{\gm r}{ \left(\dd \right)^{3/2}} ,
\label{Eulerrad}
\end{equation}
where $\rhog$ is given by Eq.~(\ref{eqhydroint}). Thus:
\begin{equation}
\rhog \left(r,z\right) = \frac{\bp\left(r\right)}{\cs^{2}\left(r,z\right)} \mathrm{e}^{-I_{1}\left(r,z\right)} ,
\label{densityf}
\end{equation}
with:
\begin{equation}
\left\lbrace 
\begin{array}{rcl}
I_{1}\left(r,z\right) & = & \displaystyle \int_{0}^{z} \gm \cs^{-2}\left(r,z'\right) \diffzp f\left(r,z'\right) \mathrm{d}z' ,\\[2ex]
f\left(r,z\right) & = & \displaystyle \frac{1}{r} - \frac{1}{\sqrt{r^{2}+z^{2}}} .
\end{array}
\right.
\label{defI}
\end{equation}
To simplify Eq.~(\ref{Eulerrad}), we first use the following identity:
\begin{equation}
\frac{1}{\rhog} \frac{\, \partial P}{\partial r} = \cs^2 \frac{\partial \, \mathrm{ln}\left(\cs^{2}\rhog\right)}{\partial r} ,
\label{simpeqhydro}
\end{equation}
which becomes with Eq.~(\ref{densityf}):
\begin{equation}
\frac{1}{\rhog} \frac{\partial P}{\partial r} = \cs^2 \frac{\mathrm{d} \, \mathrm{ln}\left(\bp \right)}{\mathrm{d} r} - \cs^{2} \frac{\partial \, I_{1}}{\partial r} .
\label{devgradp}
\end{equation}
Noting that $f\left( r,z=0\right) = 0$ and integrating $I_{1}$ by parts provides:
\begin{equation}
I_{1}\left(r,z\right) = \gm \cs^{-2} f\left(r,z \right) - \int_{0}^{z} \gm f\left(r,z'\right)\diffzp \cs^{-2}\left(r,z'\right) \mathrm{d}z' , 
\label{intpart}
\end{equation}
and:
\begin{eqnarray}
\frac{1}{\rhog} \frac{\mathrm{d}P}{\mathrm{d}r} & = & \cs^2 \frac{\mathrm{d}\, \mathrm{ln}\left(\bp\right)}{\mathrm{d}r} - \left(-\frac{\gm}{r^{2}} + \frac{\gm r}{\left(\dd \right)^{3/2}} \right.\\ \nonumber
& & \left. + \cs^{2} \gm f \diffr \cs^{-2} -\cs^{2} \frac{\partial}{\partial r}\int_{0}^{z} \gm f \diffzp \cs^{-2} \mathrm{d}z' \right) . \\ \nonumber
\label{devfingradp}
\end{eqnarray}
Then, Eq.~(\ref{Eulerrad}) becomes:
\begin{equation}
\frac{\vgtheta^{2}}{r} = \cs^2 \frac{\mathrm{d}\, \mathrm{ln}\left(\bp\right)}{\mathrm{d}r} + \frac{\gm}{r^{2}} +\cs^{2} \frac{\partial}{\partial r}\int_{0}^{z} \gm f \diffzp \cs^{-2} \mathrm{d}z' - \cs^{2} \gm f \diffr \cs^{-2} .
\label{simpvtheta}
\end{equation}
Noting that :
\begin{equation}
\frac{\partial}{\partial r} \int_{0}^{z} f \diffzp \cs^{-2} \mathrm{d}z' = \int_{0}^{z} \frac{\partial f}{\partial r} \diffzp \cs^{-2} \mathrm{d}z' + \int_{0}^{z} f \diffr \diffzp \cs^{-2} \mathrm{d}z' ,
\label{diffint}
\end{equation}
and integrating the last term of the right hand side of Eq.~(\ref{diffint}) by parts provides:
\begin{equation}
\int_{0}^{z} f \diffr \diffzp \cs^{-2} \mathrm{d}z' = f\diffr \cs^{-2} - \int_{0}^{z} \diffzp f \diffr \cs^{-2} \mathrm{d}z'.
\label{intparttrick}
\end{equation}
Therefore, Eq.~(\ref{simpvtheta}) reduces to:
\begin{equation}
\frac{\vgtheta^{2}}{r} = \frac{\gm}{r^{2}} + \cs^{2} \frac{\mathrm{d}\, \mathrm{ln} \bp}{\mathrm{d}r} + \gm \cs^{2} \int_{0}^{z} \left(\diffr f \diffzp \cs^{-2} - \diffzp f \diffr \cs^{-2} \right) \mathrm{d}z' ,
\label{vthetatrick}
\end{equation}
which can be more elegantly written as:
\begin{equation}
\frac{\vgtheta^{2}}{r} = \frac{\gm}{r^{2}} + \cs^{2} \frac{\mathrm{d}\, \mathrm{ln} \bp}{\mathrm{d}r} + \gm \cs^{2} \int_{0}^{z} \left[\nabla f \times \nabla \cs^{-2}\right].\textbf{e}_{\theta} \mathrm{d}z' .
\label{vthetaelegant}
\end{equation}
Thus, the expression of the azimuthal velocity of such a disc can be separated in three terms, called the Keplerian, the pressure gradient and the baroclinic terms respectively. This last term is neglected in most studies. For a three dimensional disc, this term rigorously cancels for $\cs=$ constant. In this case, the flow is inviscid and derives from a potential, and isobars and isodensity surfaces coincide: thus, there is no source of vorticity and the azimuthal velocity depends only on the radial coordinate. This terms also cancels out for flat discs in two dimensions. If the disc is vertically isothermal, Eq.~(\ref{vthetaelegant}) becomes:
\begin{equation}
\frac{\vgtheta^{2}}{r} = \frac{\gm}{r^{2}} + \cs^{2} \frac{\mathrm{d}\, \mathrm{ln} \bp}{\mathrm{d}r} - \gm \left[\frac{1}{r} - \frac{1}{\sqrt{r^{2} + z^{2}}} \right] \diffr \mathrm{ln}\cs^{-2}   .
\label{vthetavertiso}
\end{equation}

\subsection{Radial profiles of surface density and temperature}

In this section, we consider that the disc surface density and the temperature (and thus the sound speed) depend only on the radial coordinate and are given by the following power-law profiles:
\begin{equation}
\left\lbrace
\begin{array}{l}
\Sigma \left( r \right) =  \Sigmaz \left(\frac{r}{\Rz} \right)^{-p}  =  \Sigmap r^{-p} ,\\
T\left( r \right) =  \Tze \left(\frac{r}{\Rz} \right)^{-q}  =  \Tp r^{-q} ,\\
\cs \left( r \right)  =  \csz \left(\frac{r}{\Rz} \right)^{-q/2}  =  \csp r^{-q/2} .
\end{array}
\right.
\label{powerprofiles}
\end{equation}
For vertically isothermal thin discs, the vertical density is therefore given by Eq.~(\ref{simpangles}) with the scale height given by Eq.~(\ref{Hscale}), which can be expressed as:
\begin{equation}
H\left( r \right) = \Hz \left(\frac{r}{\Rz} \right)^{\frac{3}{2} - \frac{q}{2}} = \Hp r^{\frac{3}{2} - \frac{q}{2}} ,
\label{dvpH}
\end{equation}
with:
\begin{equation}
\Hp = \frac{\csp}{\sqrt{\gm}} .
\label{exprhp}
\end{equation}
The expression of $\rhog$ compatible with the vertical hydrostatic equilibrium and providing the power-law profile set by Eq.~(\ref{powerprofiles}) is written as:
\begin{equation}
\rhog = \rhogp r^{-x} \mathrm{e}^{-\frac{z^2}{2 H^{2}\left(r \right)}} .
\label{setrho}
\end{equation}
Indeed:
\begin{eqnarray}
\int_{-\infty}^{+\infty}\rho_\mathrm{g0}' r^{-x} \mathrm{e}^{-\frac{z^2}{2 H^{2}\left(r \right)}}\mathrm{d}z & = & \rhogp \sqrt{2\pi}H\left( r \right) r^{-x} ,\\ \nonumber
& = & \Sigmap r^{-p} .\\ \nonumber
\label{intrho}
\end{eqnarray}
Hence, with $\Sigmap = \sqrt{2 \pi}\rho_{\rm{g}0} 'H_0'$ and $x = p - \frac{q}{2} + \frac{3}{2} $,
\begin{equation}
\rhog = \frac{\Sigmap}{\sqrt{2\pi}H_0'} r^{-\left( p - \frac{q}{2} + \frac{3}{2} \right)} \mathrm{e}^{-\left[\frac{z^{2}}{2\Hp^{2}r^{3-q}} \right]} ,
\label{rhogpower}
\end{equation}
which gives the correct surface density profile when integrated with respect to $z$. With this expression of $\rhog$, $\bp \left(r\right)$ is given by:
\begin{equation}
\bp \left( r \right) = \cs^{2}\left( r \right) \rhog\left(r,z=0\right) = c_{\mathrm{s}0}'^{2} \frac{\Sigmap}{\sqrt{2\pi}H_0'} r^{-\left(p + \frac{q}{2} + \frac{3}{2} \right)}, 
\label{exprPb}
\end{equation}
which ensures that:
\begin{equation}
\frac{\mathrm{d} \, \mathrm{ln}\bp}{\mathrm{d} r} = -\frac{\left(p + \frac{q}{2} + \frac{3}{2} \right)}{r} ,
\label{logPb}
\end{equation}
and, using Eq.~(\ref{vthetavertiso}), we find:
\begin{equation}
\vgtheta = \sqrt{\frac{\gm}{r} - \left(p + \frac{q}{2} + \frac{3}{2} \right) c_{\mathrm{s}}'^{2} r^{-q} - \gm q \left(\frac{1}{r} - \frac{1}{\sqrt{r^{2} + z^{2}}} \right) } .
\label{vthetaazifull}
\end{equation}
%

\section{Dimensionless quantities and equations of motion}
\label{App:Dimensionless}

To highlight the important physical parameters involved, we set $\vkz = \sqrt{\frac{\gm}{\Rz}}$ and introduce dimensionless quantities given by the following expressions:
\begin{equation}
\left\lbrace
\begin{array}{rcl}
r/\Rz & = & R,\\
\mathcal{T}/\mathcal{T}_{0} & = & R^{-q},\\
\Sigma / \Sigmaz & = & R^{-p},\\
\vk / \vkz & = & R^{-\frac{1}{2}},\\
H / \Hz & = & R^{\frac{3}{2} - \frac{q}{2}},\\
z /\Hz & = & Z,\\
\vgtheta / \vkz & = & \dst\sqrt{\frac{1}{R} - \etaz R^{-q} - q\left(\frac{1}{R} - \frac{1}{\sqrt{R^{2}+\phiz Z^{2}}} \right)},\\
\rhog / \rhogz & = & R^{-\left(p - \frac{q}{2} + \frac{3}{2} \right)} \mathrm{e}^{-\frac{Z^{2}}{2 R^{3-q}}}.\\
\end{array}
\right.
\label{dimensionlessgas}
\end{equation}
with:
\begin{equation}
\etaz = \left(p + \frac{q}{2} + \frac{3}{2} \right)\left(\frac{\csz}{\vkz} \right)^{2} .
\label{defetaz}
\end{equation}
The dimensionless parameter $\etaz$ gives the order of magnitude of the relative discrepancy between the Keplerian motion and the gas azimuthal velocity. We note that:
\begin{equation}
\left. \frac{r \cs ^{2}}{\vk} \frac{\mathrm{d} \, \mathrm{ln}\bp}{\mathrm{d} r} \right/ \vkz = -\etaz R^{-q+1/2} .
\label{eqjustufyW77}
\end{equation}
Then, we set $\tkz = \sqrt{\frac{\Rz ^{3}}{\gm}}$ and define
\begin{equation}
\left\lbrace
\begin{array}{rcl}
\dst \frac{t}{\tkz} & = & \dst T\\[2ex]
\dst \frac{\vr}{\vkz} & = & \dst  \frac{\mathrm{d}R}{\mathrm{d}T}= \tvr\\[2ex]
\dst \frac{\vtheta}{\vkz} & = & \dst R \frac{\mathrm{d}\theta}{\mathrm{d}T} =\tvtheta\\[2ex]
\dst \frac{z}{\Hz} & = & \dst Z\\[2ex]
\dst \frac{\vz}{\Hz / \tkz} & = & \dst \frac{\mathrm{d}Z}{\mathrm{d}T}.
\end{array}
\right.
\label{dimensionlessdust}
\end{equation}
Writing the coefficient $\tilde{\mathcal{C}}\left(r,z \right)$ of the drag force of Eq. (\ref{eq:deffdrag}) as:
\begin{equation}
\tilde{\mathcal{C}}\left(r,z \right) = \mathcal{C}_{0}  \mathcal{C}\left(R,Z \right) ,
\end{equation}
and using dimensionless coordinates, we have:
\begin{equation}
\frac{ \textbf{F}_{\mathrm{D}} / m_{\mathrm{d}} }{ \vkz /  \tkz} = - \frac{\mathcal{C}\left(R,Z \right)}{\left[ \frac{s}{\left(\vkz^{\lambda} \tkz \mathcal{C}_{0} \right)^{\frac{1}{y}}} \right] ^{y}}  | \tilde{\textbf{v}}  - \tilde{\textbf{v}}_{\mathrm{g}} |^{\lambda} \left(\tilde{\textbf{v}}  - \tilde{\textbf{v}}_{\mathrm{g}} \right) .
\label{adim_drag_St}
\end{equation}
We also introduce:
\begin{equation}
s_{\mathrm{opt},0}  =  \left(\vkz^{\lambda} \tkz \mathcal{C}_{0} \right)^{\frac{1}{y}} , 
\end{equation}
and
\begin{equation}
\sz  =  \frac{s}{s_{\mathrm{opt},0}} ,
\end{equation}
so that Eq. (\ref{adim_drag_St}) becomes:
\begin{equation}
\frac{ \textbf{F}_{\mathrm{D}} / m_{\mathrm{d}} }{ \vkz /  \tkz} = - \frac{\mathcal{C}\left(R,Z \right)}{\sz ^{y}}  | \tilde{\textbf{v}}  - \tilde{\textbf{v}}_{\mathrm{g}} |^{\lambda} \left(\tilde{\textbf{v}}  - \tilde{\textbf{v}}_{\mathrm{g}} \right) .
\label{adim_drag_St2}
\end{equation}
Physically, $s_{\mathrm{opt},0}$ corresponds to the grain size at which the drag stopping time equals to the Keplerian time at $r_{0}$. In Table~\ref{tabdrag}, we give the expressions of $y$, $\lambda$, $s_{\mathrm{opt},0}$, $\mathcal{C}\left(R,Z \right)$ for the Epstein and the three Stokes drag regimes. The dimensionless equations of motion for a dust grain are then:
\begin{equation}
\left\lbrace 
\begin{array}{rcl}
\dst \frac{\mathrm{d} \tvr}{\mathrm{d} T} - \frac{\tvtheta^{2}}{R} + \frac{\tvr}{\sz ^{y}} \mathcal{C}\left(R,Z \right)  | \tilde{\textbf{v}}  - \tilde{\textbf{v}}_{\mathrm{g}} |^{\lambda}   +\frac{R}{\left(R^2 + \phiz Z^{2} \right)^{3/2}} & = & 0 \\[2ex]
\multicolumn{1}{l}{\dst \frac{\mathrm{d} \tvtheta}{\mathrm{d} T} + \frac{\tvtheta\tvr}{R} +} &&\\
\frac{\left(\tvtheta - \sqrt{\frac{1}{R} - \etaz R^{-q} - q\left(\frac{1}{R} - \frac{1}{\sqrt{R^{2}+\phiz Z^{2}}} \right)}\right)}{\sz^{y}}  \mathcal{C}\left(R,Z \right)  | \tilde{\textbf{v}}  - \tilde{\textbf{v}}_{\mathrm{g}} |^{\lambda} & = & 0 \\[2ex]
\dst \frac{\mathrm{d}^{2}Z}{\mathrm{d}T^{2}} + \frac{1}{\sz ^{y}}\frac{\mathrm{d}Z}{\mathrm{d}T} \mathcal{C}\left(R,Z \right)  | \tilde{\textbf{v}}  - \tilde{\textbf{v}}_{\mathrm{g}} |^{\lambda} + \frac{Z}{\left(R^{2} + \phiz Z^{2} \right)^{3/2}} & = & 0 .
\end{array}
\right.
\label{dustgene3d}
\end{equation}
\begin{table}
\caption{Expressions of the coefficients $y$, $\lambda$, $s_{\mathrm{opt},0}$ and $\mathcal{C}\left(R,Z \right)$ for different drag regimes.}
\label{tabdrag}
\begin{center}
\begin{tabular}{@{}lcccc@{}}
\hline Drag regime & $y$ &  $\lambda$ & $s_{\mathrm{opt},0}$ & $\mathcal{C}\left(R,Z \right)$ \\ \hline
Epstein & 1 & 0 & $\dst \frac{\Sigmaz}{\sqrt{2 \pi} \rhod}$ & $\dst R^{-\left( p +\frac{3}{2} \right)}e^{-\frac{Z^{2}}{2 R^{3-q} } } $   \\[2em]
\begin{minipage}{6em}Stokes\\ ($R_{\mathrm{g}} < 1$)\end{minipage} & 2 & 0 & $\dst \sqrt{\frac{9 \tkz \mu_{0}}{2\rhod}}$ & $\dst R^{-\frac{q}{2}}$ \\[2em]
\begin{minipage}{6.1em}Stokes\\ ($1 < R_{\mathrm{g}} < 800$)\end{minipage} & 1.6 & 0.4 & $\dst \left(\frac{18 \Rz \nu_{0}^{0.6} \Sigmaz }{v_{\mathrm{k}0}^{0.6} \sqrt{2\pi}2^{1.6}\rhod H_{0}}\right)^{\frac{1}{1.6}}$ & $\dst R^{-\left(\frac{2p}{5} + \frac{q}{10} +\frac{3}{5} \right)}   e^{-\frac{2}{5}\frac{Z^{2}}{2 R^{3-q} } }$ \\[2em]
\begin{minipage}{6em}Stokes\\ ($800 < R_{\mathrm{g}}$)\end{minipage} & 1 & 1 & $\dst \frac{ 1.32 \Rz \Sigmaz}{8 \rhod \sqrt{2\pi} H_{0}}$ & $\dst R^{-\left(p-\frac{q}{2} + \frac{3}{2} \right)}e^{-\frac{Z^{2}}{2 R^{3-q} } }$  \\[2em]
\hline
\end{tabular}
\end{center}
\end{table}
%
\section{Lemma for the different expansions}
\label{App:Lemma}

\underline{Lemma:}
Let $x$ be either $r$ or $\theta$ and $i$ the order of the perturbative expansion. If:
\begin{itemize}
\item $\tvrz = 0$, and
\item $\tilde{v}_{xi}$ can be written as a function of $R$ ($\tilde{v}_{xi} = f\left( R \right)$) with $f = \mathcal{O}\left(1\right)$ of the expansion in $\etaz$,
\end{itemize}
then, $\dst \frac{\mathrm{d}\tilde{v}_{xi} }{\mathrm{d} T} $ is of order $\mathcal{O}\left(\etaz\right)$.\\
\noindent\underline{Proof}:
\begin{equation}
\frac{\mathrm{d} \tilde{v}_{xi} }{\mathrm{d} T} = \frac{\mathrm{d} \tilde{v}_{xi} }{\mathrm{d} R} \frac{\mathrm{d}R}{\mathrm{d}T} = \tvr f'\left(R\right) = \etaz \tvru f'\left(R\right) + \mathcal{O}\left(\etazsq \right) = \mathcal{O}\left(\etaz \right).
\label{lemmeproof}
\end{equation}
%
\section{Epstein regime: perturbation analysis}
\label{App:Epstein_perturb}

\begin{itemize}
\item \underline{Order $\mathcal{O}\left(1\right)$:}
At this order of expansion, $\etaz R^{-q}$ is negligible compared to $\frac{1}{R}$. Thus, substituting Eq.~(\ref{dvptnaka}) into Eq.~(\ref{radialseulmodif}) provides
\begin{equation}
\left\lbrace
\begin{array}{l}
\dst \frac{\mathrm{d} \tvrz}{\mathrm{d} T} = \dst \frac{\tvthetaz^{2}}{R} - \frac{\tvrz}{\sz}R^{-\left(p+\frac{3}{2} \right)} -\frac{1}{R^2} \\
\dst \frac{\mathrm{d} \tvthetaz}{\mathrm{d} T} = \dst -\frac{\tvthetaz \tvrz}{R} - \frac{\left(\tvthetaz - \sqrt{\frac{1}{R}}\right)}{\sz}R^{-\left(p+\frac{3}{2} \right)} .
\end{array}
\right.
\label{devnakaz}
\end{equation}
At this stage, we do not know the order of $\frac{\mathrm{d}\tvthetaz}{\mathrm{d} T}$. We show in the lemma of Appendix~\ref{App:Lemma} that $\frac{\mathrm{d} \tvthetaz}{\mathrm{d} T} = \mathcal{O}\left(\etaz \right)$. Applying this lemma, we see that at the order $\mathcal{O}\left(1 \right)$, taking $\tvrz = 0$ and $\tvthetaz = \sqrt{\frac{1}{R}}$ (which ensures that $\frac{\mathrm{d} \tvthetaz }{\mathrm{d} T}  = \mathcal{O}\left(\etaz \right)$) is a relevant solution for the equations of motion (which corresponds to circular Keplerian motion). Thus,
\begin{equation}
\left\lbrace
\begin{array}{l}
\dst \tvrz = \dst 0  \\
\dst \tvthetaz = \dst \sqrt{\frac{1}{R}} .
\end{array}
\right.
\label{devnakaordrezero}
\end{equation}
\item \underline{Order $\mathcal{O}\left(\etaz\right)$:}
Applying the lemma in this order of expansion and noting that
\begin{equation}
\frac{\mathrm{d} \tvthetaz }{\mathrm{d} T} = - \frac{1}{2} R^{-3/2} \tvr =  -\frac{\etaz}{2}R^{-3/2}\tvru + \mathcal{O}\left(\etaz^{2} \right),
\label{devdvthetazdt}
\end{equation}
Eq.~(\ref{radialseulmodif}) becomes
\begin{equation}
\left\lbrace
\begin{array}{l}
\dst 0  =  - \frac{R^{-\left(p+\frac{3}{2} \right)}}{\sz}\tvru + \dst 2 R^{-3/2}\tvthetau \\
\dst 0  =  \dst -\frac{1}{2}R^{-3/2}\tvru - \frac{\left(\tvthetau - \frac{1}{\etaz}\left( \sqrt{\frac{1}{R} - \etaz R^{-q}} - \sqrt{\frac{1}{R}}  \right)\right)}{\sz}R^{-\left(p+\frac{3}{2} \right)} .
\end{array}
\right.
\label{devnakau}
\end{equation}
Solving the linear system Eq.~(\ref{devnakau}) for $\left(\tvru,\tvthetau \right)$ provides
\begin{equation}
\left\lbrace
\begin{array}{l}
\dst \tvru = \dst - \frac{1}{\etaz} \frac{2 \sz R^{p- \frac{1}{2} } \left(1 - \sqrt{1 - \etaz R^{-q+1}} \right)}{1 + R^{2p} \szsq}\\
\dst \tvthetau = \dst - \frac{1}{\etaz} \frac{R^{-\frac{1}{2}} \left(1 - \sqrt{1 - \etaz R^{-q+1}} \right)}{1 + R^{2p}\szsq}.
\end{array}
\right.
\label{directsolsyst}
\end{equation}
\end{itemize}
In addition to the expression of $\tvr$ given in Sect.~\ref{Sec:RadialMotion}, we also note that
\begin{equation}
\begin{array}{rcl}
\tvtheta & = & \dst \sqrt{\frac{1}{R}} + \etaz \tvthetau + \mathcal{O}\left(\etazsq \right) \\
& = & \dst \sqrt{\frac{1}{R}} - \frac{R^{-\frac{1}{2}} \left(1 - \sqrt{1 - \etaz R^{-q+1}} \right)}{1 + R^{2p}\szsq} + \mathcal{O}\left(\etazsq \right),
\end{array}
\label{nakatvrtehtau}
\end{equation}
which provides with Eq.~(\ref{approxsmalletaz})
\begin{equation}
\tvtheta = \sqrt{\frac{1}{R}} - \frac{\etaz R^{-q+\frac{1}{2}}}{2\left(1 + R^{2p}\szsq \right)} .
\label{nakatvthetausimp}
\end{equation}
%
%
\section{Link with \citetalias{Weidendust1977}'s original derivation}
\label{App:W77}

Following \citetalias{Weidendust1977}'s historic reasoning for small grains (see Sect.~\ref{Sec:Amode}), we perform a perturbative expansion of the radial equation of motion with the $\sz$ variable. We verify that taking the limit at small $\etaz$ provides the expression found for the A-mode in the \citetalias{Nakagawa1986} expansion. (Formally, we will show that $\lim\limits_{\substack{\sz \ll 1}} \lim\limits_{\substack{\etaz \ll 1}}\left[... \right] = \lim\limits_{\substack{\etaz \ll 1}} \lim\limits_{\substack{\sz \ll 1}}\left[... \right]$ ). Hence, we set:
\begin{equation}
\left\lbrace
\begin{array}{l}
\dst \tvr  =  \dst \tvrz + \sz \tvru + \szsq \tvrd + \mathcal{O}\left(\sz^{3}\right),\\
\dst \tvtheta  =  \dst \tvthetaz + \sz \tvthetau + \szsq \tvthetad + \mathcal{O}\left(\sz^{3}\right),
\end{array}
\right.
\label{dvptW77small}
\end{equation}
where we have used for convenience the same formalism as for the expansion in $\etaz$ -- see Eq.~(\ref{dvptnaka}) (noting of course that $\tilde{v}$ represents different functions). An important point is that the lemma of Appendix \ref{App:Lemma} holds when substituting $\sz$ to $\etaz$. Therefore, substituting Eq.~(\ref{dvptW77small}) into Eq.~(\ref{radialseulmodif}) provides the equations of motion for different orders of $\mathcal{O}\left(\sz \right)$:
\begin{itemize}
\item \underline{Order $\mathcal{O}\left(\frac{1}{\sz}\right)$:} Eq.~(\ref{radialseulmodif}) provides $\mathcal{O}\left(1\right)$ expressions for the velocities:
\begin{equation}
\left\lbrace
\begin{array}{r@{\ }l}
\dst \tvrz = & \dst 0,\\
\dst \tvthetaz = & \dst \sqrtexpr .
\end{array}
\right.
\label{dvptW77smallmu}
\end{equation}
In this order of expansion, the azimuthal velocity corresponds to the sub-Keplerian velocity of the gas. There is no radial motion.

\item \underline{Order $\mathcal{O}\left(1 \right)$:}
\begin{equation}
\left\lbrace
\begin{array}{r@{\ }l}
\dst \tvru = & \dst -2 R^{p-\frac{1}{2}} \left(1 - \sqrt{1 - \etaz R^{-q+1}} \right),\\
\dst \tvthetau = & \dst 0.
\end{array}
\right.
\label{dvptW77smallz}
\end{equation}

\item \underline{Order $\mathcal{O}\left(\sz \right)$:}
\begin{equation}
\left\lbrace
\begin{array}{r@{\ }l}
\dst \tvrd = & \dst 0,\\
\dst \tvthetad = & \dst R^{p+\frac{3}{2}}\left[\etaz \sqrtexpr R^{p-q-\frac{1}{2}}\right. \\
& \qquad\quad\left.\dst +\frac{1}{2}\frac{-\frac{1}{R^{2}} + \etaz q R^{-q-1}}{\sqrtexpr \etaz R^{p-q+\frac{1}{2}}} \right] .
\end{array}
\right.
\label{dvptW77smallu}
\end{equation}
\end{itemize}
Finally, we obtain expressions for $\tvr$ and $\tvtheta$:
\begin{equation}
\left\lbrace
\begin{array}{r@{\ }l}
\dst \tvr = & \dst -2\sz R^{p-\frac{1}{2}} \left(1 - \sqrt{1 - \etaz R^{-q+1}} \right) + \mathcal{O}\left(S_{0}^{3}\right),\\[2ex]
\dst \tvtheta = & \dst \sqrtexpr + \szsq R^{p+\frac{3}{2}}\left[\etaz \sqrtexpr R^{p-q-\frac{1}{2}} \right. \\
& \qquad\quad\left.\dst +\frac{1}{2}\frac{-\frac{1}{R^{2}} + \etaz q R^{-q-1}}{\sqrtexpr \etaz R^{p-q+\frac{1}{2}}} \right] + \mathcal{O}\left(S_{0}^{3}\right) .
\end{array}
\right.
\label{dvptW77smalld}
\end{equation}

We now compare the \citetalias{Nakagawa1986} expansion at $R^{p} \sz \ll 1$ (A-mode) and the \citetalias{Weidendust1977} expansion at $\etaz \ll 1$.
\begin{itemize}
\item \underline{\citetalias{Nakagawa1986}:} From Eqs.~(\ref{nakatvrusimp}) and Eq.~(\ref{nakatvthetausimp}):
\begin{equation}
\left\lbrace
\begin{array}{r@{\ }l}
\dst \tvr = & \dst - \frac{\etaz \sz R^{p-q+\frac{1}{2}}}{1+R^{2p}\szsq} + \mathcal{O}\left(\etaz ^{2} \right)\\
= & \dst - \etaz \sz R^{p-q+\frac{1}{2}} + \mathcal{O}\left(\etaz ^{2}\right) + \mathcal{O}\left(\sz ^{2}\right),\\[2ex]
\dst \tvtheta = & \dst \sqrt{\frac{1}{R}} - \frac{\etaz}{2} \frac{R^{-q+\frac{1}{2}}}{1 + \szsq R^{2p}} + \mathcal{O}\left(\etaz ^{2} \right)\\
= & \dst \sqrt{\frac{1}{R}} - \frac{\etaz}{2}R^{-q+\frac{1}{2}} + \frac{\etaz \szsq}{2} R^{2p-q+\frac{1}{2}} + \mathcal{O}\left(\etaz ^{2}\right) + \mathcal{O}\left(\sz ^{3} \right).
\end{array}
\right.
\label{NSH86radazim}
\end{equation}
\item \underline{\citetalias{Weidendust1977} small grains:} From Eq.~(\ref{dvptW77smalld}):
\begin{equation}
\left\lbrace
\begin{array}{r@{\ }l}
\dst \tvr = &\dst  -2\sz R^{p-\frac{1}{2}} \left(1 - \sqrt{1 - \etaz R^{-q+1}} \right) + \mathcal{O}\left(\sz ^{2}\right)\\
= & \dst - \etaz \sz R^{p-q+\frac{1}{2}} + \mathcal{O}\left(\sz ^{2}\right) + \mathcal{O}\left(\etaz ^{2}\right),\\[2ex]
\dst \tvtheta = & \dst \sqrtexpr + \szsq R^{p+\frac{3}{2}} \Big[\etaz \sqrtexpr R^{p-q-\frac{1}{2}}\\
& \dst + \frac{1}{2}\frac{\etaz R^{p-q+\frac{1}{2}} \left(-\frac{1}{R^{2}}+ \etaz q R^{-q-1} \right)}{\sqrtexpr} \Big] + \mathcal{O}\left(\sz ^{3}\right)\\
= & \dst \sqrt{\frac{1}{R}} - \frac{\etaz}{2}R^{-q+\frac{1}{2}} + \frac{\etaz \szsq}{2} R^{2p-q+\frac{1}{2}} + \mathcal{O}\left(\sz ^{3}\right) + \mathcal{O}\left(\etaz ^{2}\right) .
\end{array}
\right.
\label{W77smallexprad}
\end{equation}
\end{itemize}
Clearly, Eqs.~(\ref{NSH86radazim}) and (\ref{W77smallexprad}) are identical, demonstrating that the theories of \citetalias{Weidendust1977} and \citetalias{Nakagawa1986} are consistent. We also note that if the simplification of Eq.~(\ref{approxsmalletaz}) is not performed, the two \citetalias{Weidendust1977} expansions directly appear as the expansion of \citetalias{Nakagawa1986} in $\mathcal{O}\left(\sz\right)$ or $\mathcal{O}\left(\sz ^{-1}\right)$.

Now, in the case of large grains, we perform a perturbative expansion of the radial equation of motion with respect to $\frac{1}{\sz}$ while assuming that $\sz R^{p} \gg 1$, and verify that taking the limit at small $\etaz$ provides the expression found for the B-mode in the \citetalias{Nakagawa1986} expansion.

Taking the same precautions as for the previous expansions, we write:
\begin{equation}
\left\lbrace
\begin{array}{r@{\ }l}
\dst \tvr = & \dst \tvrz + \frac{1}{\sz} \tvru +  \mathcal{O}\left(\frac{1}{\sz^{2}}\right),\\
\dst \tvtheta = & \dst \tvthetaz + \frac{1}{\sz} \tvthetau + \mathcal{O}\left(\frac{1}{\sz^{2}}\right).
\end{array}
\right.
\label{dvptW77large}
\end{equation}
Following the same method as for the small grain sizes expansion, we obtain:

\begin{itemize}
\item \underline{Order $\mathcal{O}\left(1\right)$:}
\begin{equation}
\left\lbrace
\begin{array}{l}
\tvrz  =  0,\\
\tvthetaz  = \dst \sqrt{\frac{1}{R}} .
\end{array}
\right.
\label{dvptW77largez}
\end{equation}
It this order of expansion, the azimuthal velocity of the grain is the standard Keplerian velocity.

\item \underline{Order $\mathcal{O}\left(\frac{1}{\sz} \right)$:}
\begin{equation}
\left\lbrace
\begin{array}{l}
\tvru  = \dst -2\left(\sqrt{\frac{1}{R}} - \sqrtexpr \right)R^{-p},\\
\tvthetau  =  0.
\end{array}
\right.
\label{dvptW77largeu}
\end{equation}
\end{itemize}
The expansion at higher order is more complicated and will not be used for further developments. At the order $\mathcal{O}\left(\frac{1}{S_{0}^{2}}\right)$, we have for $\tvr$ and $\tvtheta$:
\begin{equation}
\left\lbrace
\begin{array}{l}
\dst \tvr  =  \dst -\frac{2}{\sz}\left(\sqrt{\frac{1}{R}} - \sqrtexpr \right)R^{-p} + \mathcal{O}\left(\frac{1}{S_{0}^{2}}\right),\\
\dst \tvtheta  =  \dst \sqrt{\frac{1}{R}} + \mathcal{O}\left(\frac{1}{S_{0}^{2}}\right) .
\end{array}
\right.
\label{dvptW77largefin}
\end{equation}
We now compare the expressions provided by the \citetalias{Nakagawa1986} expansion at $R^{2p} \szsq \gg 1$ (B-mode) and the \citetalias{Weidendust1977} expansion at $\etaz \ll 1$ for the radial velocity:
\begin{itemize}

\item \underline{\citetalias{Nakagawa1986}:}
\begin{eqnarray}
\tvr & = & - \frac{\etaz \sz R^{p-q+\frac{1}{2}}}{1+R^{2p}\szsq} + \mathcal{O}\left(\etaz ^{2} \right),\\ \nonumber
& = & - \frac{\etaz}{\sz} R^{-p-q+\frac{1}{2}} + \mathcal{O}\left(\etaz ^{2} \right) + \mathcal{O}\left(\frac{1}{\szsq} \right).\\ \nonumber
\label{NSH86radlarge}
\end{eqnarray}
\item \underline{\citetalias{Weidendust1977} large grains:}
\begin{eqnarray}
\tvr & = & -\frac{2}{\sz} R^{-p-\frac{1}{2}} \left(1 - \sqrt{1 - \etaz R^{-q+1}} \right) + \mathcal{O}\left(\frac{1}{\szsq} \right),\\ \nonumber
& = & - \frac{\etaz}{\sz} R^{-p-q+\frac{1}{2}} + \mathcal{O}\left(\frac{1}{\szsq} \right) + \mathcal{O}\left(\etaz ^{2} \right) .\\ \nonumber
\label{W77largeexprad}
\end{eqnarray}
\end{itemize}
Once again, we show that the \citetalias{Weidendust1977} and \citetalias{Nakagawa1986} theories are consistent.

\section{Asymptotic radial behaviour of a single grain}
\label{App:Asymptotic_radial}
\begin{figure}
\resizebox{\hsize}{!}{\includegraphics[angle=-90]{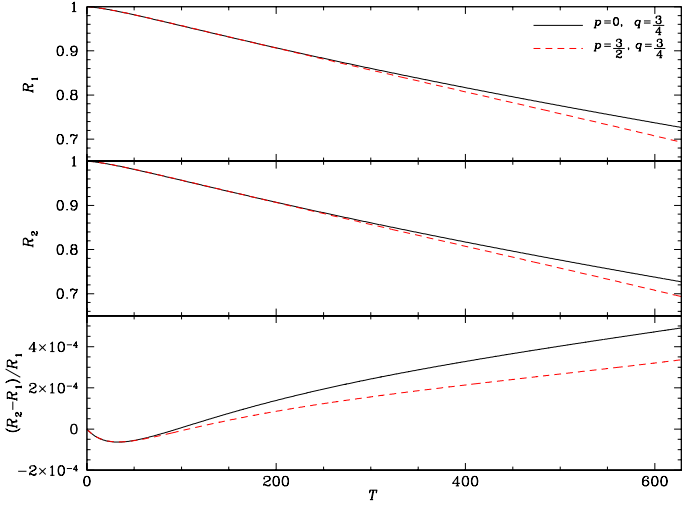}}
\caption{The discrepancy (bottom panel) between the exact motion (top panel) and its NSH86 approximation (central panel) is negligible. This is illustrated plotting the radial motion of dust grains for $\sz=10^{-2}$, $\etaz = 10^{-2}$ and for $p=0,q=\frac{3}{4}$ (solid) and $p=\frac{3}{2},q=\frac{3}{4}$ (dashed). Top: $R_{1}\left(T\right)$, middle: $R_{2}\left(T\right)$, bottom: relative difference $\left( R_{2}\left(T\right) - R_{1}\left(T\right) \right) / R_{1}\left(T\right)$.} 
\label{plotdiff}
\end{figure}
Noting $R_{1}\left(T\right)$ the position of a grain integrated directly from the equation of motion (Eq.~(\ref{radialseulmodif})) and $R_{2}\left(T\right)$ the position integrated from the \citetalias{Nakagawa1986} approximation (Eq.~(\ref{implicitNSH})), we highlight (Fig.~\ref{plotdiff}) that the discrepancy between the motion from the exact equations and its \citetalias{Nakagawa1986} approximation is negligible (the relative error is lower than $10^{-3}$ for all the considered sizes). It is therefore justified to use the analytical results derived in Sect.~\ref{Sec:RadialMotion} to interpret the grain behaviour.
\begin{figure*}
\sidecaption
\includegraphics[width=12cm]{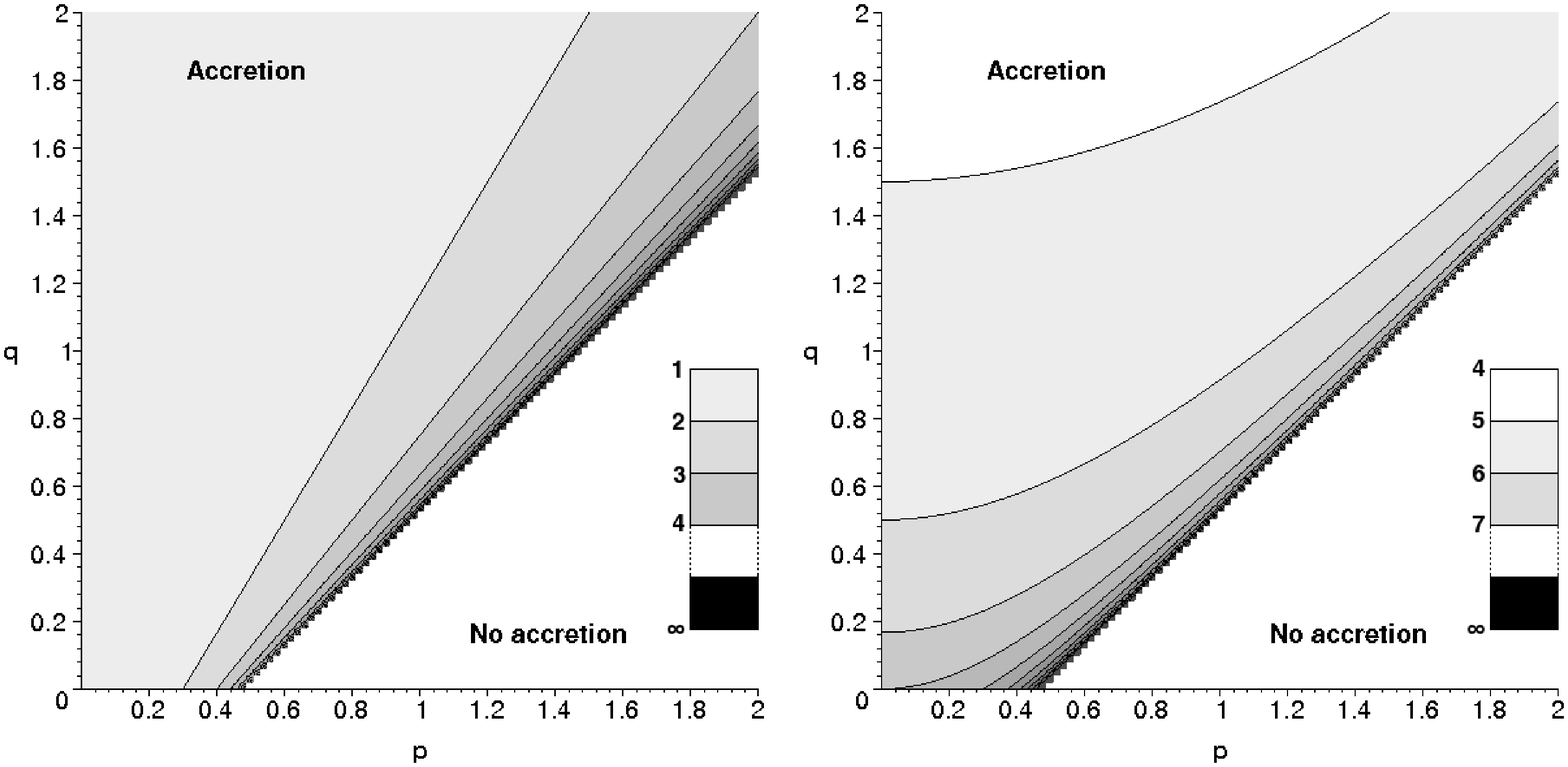}
\caption{Values of $\sm$ (left) and $\etaz \Tm\left( \sm \right)$ (right) in the ($p$,$q$) plane.}
\label{Sm-etazTmfig}
\end{figure*}
Thus, from Eq.~(\ref{implicitNSH}), we see that the time for a grain starting at $R = 1$ to reach some final radius $R_{\mathrm{f}}$ is minimized for an optimal grain size $S_{\mathrm{m,f}}$ given by
\begin{equation}
S_{\mathrm{m,f}} = \sqrt{ \frac{p+q+\frac{1}{2}}{-p+q+\frac{1}{2}} \times k_{\mathrm{f}} \left(R_{\mathrm{f}} \right)} \,,
\label{exprsmr}
\end{equation}
with
\begin{equation}
k_{\mathrm{f}} \left(R_{\mathrm{f}} \right)  =
\left\lbrace
\begin{array}{ll}
\frac{\left(1 - R_{\mathrm{f}}^{-p+q+\frac{1}{2}} \right)}{\left(1 - R_{\mathrm{f}}^{p+q+\frac{1}{2}} \right)} & \mathrm{if}~p+q+\frac{1}{2} \ne 0\\[1ex]
-\mathrm{ln}\left(R_{\mathrm{f}}\right) & \mathrm{if}~p+q+\frac{1}{2} = 0.\\
\end{array}
\right.
\label{exprdmr}
\end{equation}
As shown in Eq.~(\ref{implicitNSH}), the outcome of the grain radial motion depends on the value of $-p+q+\frac{1}{2}$:
\begin{itemize}
\item If $-p+q+\frac{1}{2} \le 0$:
\begin{equation}
\lim\limits_{\substack{T \to +\infty}} R = 0 .
\label{modeNGHinf}
\end{equation}
For such disc profiles, all grains pile-up and fall onto the central star in an infinite time. Indeed, the surface density profile given by $p\ge q + \frac{1}{2}$ is steep enough to counterbalance the increase of the acceleration due to the pressure gradient. Therefore, grains fall onto the central star in an infinite time, whatever their initial size. Such an evolution happens because the grains always end migrating in the A-mode when they reach the disc's inner regions. A crucial consequence is that grains are not depleted on the central star and therefore stay in the disc where they can potentially form planet embryos.
\item If $-p+q+\frac{1}{2} > 0$:
\begin{equation}
\lim\limits_{\substack{T \to \Tm}} R = 0 ,
\label{modeNGHfini}
\end{equation}
where
\begin{equation}
\Tm = \frac{1}{\etaz \sz} \left( \frac{1}{-p+q+\frac{1}{2}}+ \frac{\szsq}{p+q+\frac{1}{2}} \right) .
\label{defTmNGH}
\end{equation}
In this case, grains fall onto the central star in a finite time. The surface density profile given by $p<q+\frac{1}{2}$ is now too flat to counterbalance the increasing acceleration due to pressure gradient. We note that:
\begin{itemize}
\item For small sizes ($\sz \ll 1$), $\Tm = \mathcal{O}\left( \sz \etaz\right)$.
\item For large sizes ($\sz \gg 1$), $\Tm = \mathcal{O}\left(\frac{\sz}{\etaz}\right)$. 
\item $\Tm$ reaches a minimal value for a size $\sm$ given by
\begin{equation}
\sm = \sqrt{ \frac{p+q+\frac{1}{2}}{-p+q+\frac{1}{2}}} .
\label{defsm}
\end{equation}
Therefore
\begin{equation}
\Tm\left( \sm \right) = \frac{2}{\etaz \sqrt{\left(p + q + \frac{1}{2} \right)\left(-p + q + \frac{1}{2} \right) }} .
\label{deftmsm}
\end{equation}
$\sm$ is of order unity and corresponds to an optimal size of migration. Values of $\sm$ and $\etaz \Tm\left( \sm \right)$ in the $(p,q)$ plane are shown in Fig.~\ref{Sm-etazTmfig}. When $S \simeq \sm  $, both the A- and B-modes contribute in an optimal way to the grains radial motion.
\end{itemize}
In this case, grains can be efficiently accreted by the central star if $S \simeq \sm = \mathcal{O}\left(1\right)$. Thus, they can not contribute to the formation of pre-planetesimals. This process is called the radial-drift barrier for planet formation.
\end{itemize}

\end{appendix}

\end{document}